\documentclass{aa}
\usepackage[english]{babel}
\usepackage[varg]{txfonts}

\usepackage[usenames,dvipsnames]{xcolor}

\usepackage{epsfig}
\usepackage{graphicx,natbib}
\usepackage{amsmath,amssymb}
\usepackage{hyperref}
\usepackage{wasysym}
\usepackage{natbib}
\usepackage{ulem}
\usepackage{cancel}
\usepackage{subfigure}
\usepackage{gensymb}
\usepackage{stmaryrd}

\hypersetup{
    colorlinks=true,
    citecolor=blue,
    linkcolor=red,
    urlcolor=black
    }

\def\ga{\,\hbox{\hbox{$ > $}\kern -0.8em \lower 1.0ex\hbox{$\sim$}}\,}
\def\la{\,\hbox{\hbox{$ < $}\kern -0.8em \lower 1.0ex\hbox{$\sim$}}\,}
\def\aap{A\&A}

\begin{document}

\def\nat{Nature }
\def\apj{Astrophys. J. }
\def\apjs{Astrophys. J., Suppl. Ser. }
\def\apjl{Astrophys. J., Lett. }
\def\apss{Astrophys. and Space Science}

\title{Impact of the Hall effect in star formation and the issue of angular momentum conservation}
\titlerunning{The Hall effect in star formation with AMR codes}

\author{P. Marchand \inst{1,2}
       \and B. Commer\c con\inst{2}
       \and G. Chabrier \inst{2,3}
}

\institute{
Department of Earth and Space Science, Osaka University, Toyonaka, Osaka 560-0043, Japan
\and CRAL, Ecole normale sup\'erieure de Lyon, UMR CNRS 5574, Universit\'e de Lyon, France
\and School of Physics, University of Exeter, Exeter, EX4 4QL, UK
}

\authorrunning{P. Marchand et~al.}

\date{}

\abstract{We present an implementation of the Hall term in the non-ideal magnetohydrodynamics equations into the adaptive-mesh-refinement code {\ttfamily RAMSES} to study its impact on star formation. Recent works show that the Hall effect heavily influences the regulation of the angular momentum in collapsing dense cores, strengthening or weakening the magnetic braking. Our method consists of a modification of the two-dimensional constrained transport scheme. Our scheme shows convergence of second-order in space and the frequency of the propagation of whistler waves is accurate. We confirm previous results, namely that during the collapse, the Hall effect generates a rotation of the fluid with a direction in the mid-plane that depends on the sign of the Hall resistivity, while counter-rotating envelopes develop on each side of the mid-plane. However, we find that the predictability of our numerical results is  severely limited. The angular momentum is not conserved in any of our dense core-collapse simulations with the Hall effect: a large amount of angular momentum is generated within the first Larson core, a few hundred years after its formation, without compensation by the surrounding gas. This issue is not mentioned in previous studies and may be correlated to the formation of the accretion shock on the Larson core. We expect that this numerical effect could be a serious issue in star formation simulations.}


\maketitle

\section{Introduction}

\vspace{0.17cm}

Star formation originates in the collapse and fragmentation of molecular clouds. These giant structures of the interstellar medium harbor thousands of solar masses of gas, mostly Hydrogen ($\sim 74$\% by mass) and Helium ($\sim 25$\%). This fragmentation leads to the creation of over-dense regions, the dense cores, that dynamically decouple from the rest of the cloud. When a dense core gravitationally collapses onto itself, the accumulation of matter and the rise of the temperature form a star embryo, a protostar.

\citet{Larson1969} describes the early stages of the collapse of dense cores. Initially, the collapse is isothermal as the gas-dust mixture is transparent to its own blackbody radiation. When the density reaches $10^{-13}$ g cm$^{-3}$, this radiation is trapped and heats up the gas, leading to the formation of an hydrostatic core called the first Larson core. The core contracts adiabatically until the temperature reaches $2000$ K. At this moment, H$_2$ molecules dissociate in an endothermic reaction, thus creating a gap of energy that allows a second collapse. Once all hydrogen is atomic, the adiabatic contraction resumes in what is called the second Larson core. While the second core is considered to be a protostar, the first Larson core remains a theoretical object not confirmed by observations because of its short life-time (few thousands years at maximum). Nevertheless, serious candidates have been revealed by recent observations \citep[e.g.,][]{2012A&A...547A..54P,2015A&A...577L...2G}.

The dynamics of the collapse is affected by magnetic fields threading the dense cores. It has been well-established that they play a major role during the contraction of dense cores, and they are considered as the best way to solve the angular momentum problem. Observations show that the specific angular momentum of dense cores ($\sim 10^{21}$ cm$^2$ s$^{-1}$) is three orders of magnitude higher than those of young stars ($\sim 10^{18}$ cm$^2$ s$^{-1}$) \citep{1995ARA&A..33..199B}. The magnetic field is able to transport the angular momentum from the inner parts of the dense cores to their envelop through the magnetic braking. In the last twenty years, there have been several studies investigating the impact of the magnetic braking during protostellar collapse through numerical simulations using the approximation of ideal magnetohydrodynamics (MHD), in which the charged particles are perfectly coupled with the neutral gas. However, the braking is so efficient in removing the angular momentum that, at best, only small rotationnally supported disks could be formed \citep{AllenShuLi,2004ApJ...616..266M,GalliShuLizano2006,PriceBate2007,HennebelleTeyssier2008,HennebelleFromang2008,commercon10,DA1}. While the conditions of their formation is still matter to debate, such disks evolve in protoplanetary disks, environments of planet formation that have been widely observed \citep{2013A&A...560A.103M,2015ApJ...808L...3A,2017A&A...606A..35G,2017A&A...606L...3F}. Additionally, the flux-freezing approximation of ideal MHD causes a magnetic pile-up in the protostar, leading to values of the magnetic flux up to three orders of magnitude higher than those observed \citep{1984FCPh....9..139N,TrolandCrutcher2008,2008ASPC..384..145J}.

Among the possible solutions to solve both the angular momentum and the magnetic flux problem, non-ideal MHD is considered to be the most efficient and robust. Non-ideal MHD describes the resistance encountered by the charged particles into a not fully-ionized fluid. The first studies accounted for the Ohmic and ambipolar diffusion during the early stages of the collapse \citep[i.e.,][]{Machida_etal06,DuffinPudritz,MellonLi2009,Kunz2,2015ApJ...801..117T,DA1}. They showed that these processes could indeed temper the magnetic braking and regulate the magnetic flux. The third non-ideal MHD process, the Hall effect, has been included only recently in numerical simulations. Its resistivity value indicates that its influence on the system can be similar in strength to the ambipolar and Ohmic diffusion at densities ranging from the isothermal stage to the end of the first Larson core \citep{2016A&A...592A..18M,2016PASA...33...41W}. It is therefore necessary to quantify its impact through numerical simulations. Recent studies have shown that the Hall effect does indeed play a major role in the regulation of the angular momentum \citep{2011ApJ...733...54K, Tsukamoto2015, Wurster2015, 2017PASJ...69...95T} and in the formation of binary stars \citep{2017MNRAS.466.1788W}.

Several astrophysical codes, both Eulerian and Lagrangean, already include the Hall effect, through various numerical methods. The first star-formation simulations including the Hall term (\citet{Tsukamoto2015}, \citet{2017PASJ...69...95T}), as
well as the most recent studies 
(\citet{Wurster2015}, \citet{2017MNRAS.466.1788W}, \citet{2018MNRAS.475.1859W}) have been performed with SPH codes.
The protoplanetary disks shearing box simulations by \citet{Lesur2014} uses the grid-based code PLUTO \citep{2007ApJS..170..228M}. The Hall effect is also implemented in the AMR codes BASTRUS \citep{Toth2008} and ZEUS \citep{StoneNorman,Krasnopolsky}, as well as in the multimethods code GIZMO \citep{2015MNRAS.450...53H,2017MNRAS.466.3387H}.

The aim of this paper is first to present an original implementation of the Hall effect in the adaptive-mesh-refinement (AMR) code {\ttfamily RAMSES} and to apply it in protostellar collapse studies.
The paper is organized as follow. First, we briefly recall the basics of the theory about the Hall effect in Section \ref{sect_theory}. Second, we detail our numerical implementation in Section \ref{sect_implement} and present validation tests of our scheme in Section \ref{sect_tests}. Section \ref{sect_collapse} is devoted to the protostellar collapse models in which we incorporate the Hall effect. Finally, in Section \ref{sect_ccl} we give our conclusions.

\section{Theory} \label{sect_theory}

\subsection{Equations of non-ideal MHD}

We write the complete set of non-ideal MHD equations in conservative form with self-gravity as a source term
\begin{align}
  &\frac{\partial \rho}{\partial t} + \nabla \cdot \left[\rho \mathbf{u}\right]  = 0, \label{testmass}\\
  &\frac{\partial \rho \mathbf{u}}{\partial t} + \nabla \cdot \left[\rho \mathbf{u}\mathbf{u} + \left(P+\frac{B^2}{2}\right) \mathbb{I} - \mathbf{BB}\right]  = -\rho \nabla \Phi, \label{testmomentum} \\ 
  &\frac{\partial \mathbf{B}}{\partial t} - \nabla \times  \left[\mathbf{u} \times \mathbf{B} + \mathbf{E}_\mathrm{NIMHD} \right]  = 0,   \label{testinduction} \\
  &\frac{\partial E_\mathrm{tot}}{\partial t} + \nabla \cdot \left[ \left(E_\mathrm{tot} + P_\mathrm{tot}\right)\mathbf{u} - \left(\mathbf{u}  \cdot \mathbf{B} \right) \mathbf{B} - \mathbf{E}_\mathrm{NIMHD} \times \mathbf{B}\right] = -\rho \mathbf{u} \cdot \nabla \Phi, \label{testenergy}
\end{align}
where $\rho$ is the density, $\mathbf{u}$ the fluid velocity, $P$ the thermal pressure, $\mathbf{B}$ the magnetic field, $\Phi$ the gravitational potential. $E_\mathrm{tot}$ and $P_\mathrm{tot}$ are the total energy and total pressure
\begin{align}
  E_\mathrm{tot} &= \rho \epsilon + \frac{1}{2}\rho u^2 + \frac{1}{2} B^2, \\
  P_\mathrm{tot} &= (\gamma-1)\rho \epsilon + \frac{1}{2} B^2,
\end{align}
where $\epsilon$ is the specific internal energy and $\gamma$ the adiabatic index. $\mathbf{E}_\mathrm{NIMHD}$ represents the electric field created by the three non-ideal MHD effects, and reads \citep{nakano2002}\citep{masson_nimhd}
\begin{equation}
  \mathbf{E}_\mathrm{NIMHD} = -\eta_\Omega \mathbf{J} - \frac{\eta_\mathrm{H}}{||\mathbf{B}||} \left(\mathbf{J} \times \mathbf{B}\right) + \frac{\eta_\mathrm{AD}}{||\mathbf{B}||^2} \left(\left(\mathbf{J} \times \mathbf{B}\right)\times \mathbf{B}\right),
\end{equation}
with $\eta_\Omega$, $\eta_\mathrm{H}$ and $\eta_\mathrm{AD}$ the Ohmic, Hall, and ambipolar resistivities, and $\mathbf{J}=\nabla \times \mathbf{B}$ the electric current. The MHD equations have been written in the Gaussian cgs-units convention, such that the permeability constant $\mu_0=1$.

\subsection{The Hall effect}

We first performed a perturbation analysis to characterize the Hall effect. We ignored all forces but the Lorentz force, as well as the Ohmic and ambipolar diffusion terms.
Let us consider a fluid at rest in an equilibrium state, threaded by a uniform magnetic field $\mathbf{B}=B\mathbf{e}_z$. We introduced small perturbations $\delta u_x$, $\delta u_y$, $\delta B_x$, $\delta B_y$ of the velocity and magnetic fields in the $x$ and $y$-directions, propagating along the $z-$direction at the frequency $\omega$. Each perturbation is therefore proportional to $\mathrm{exp}(i\omega t - ikz)$. 
Considering only the Hall effect, the induction Equation \eqref{testinduction} reads
\begin{equation}
  \frac{\partial \mathbf{B}}{\partial t} =  \nabla \times \left[\mathbf{u} \times \mathbf{B} - \frac{\eta_\mathrm{H}}{||\mathbf{B}||} \mathbf{J} \times {\mathbf{B}}\right]. \label{redind}
\end{equation}
It is coupled with the momentum Equation \eqref{testmomentum}, with only the Lorentz force that we write in its non-conservative form
\begin{equation}
\rho \frac{\partial \mathbf{u}}{\partial t} = \mathbf{J} \times \mathbf{B}. \label{redmom}
\end{equation}

We introduced the perturbations in Equations \eqref{redind} and \eqref{redmom}, and cancel out the zeroth and second order terms
\begin{align}
&\rho i\omega \delta u_x = -ikB\delta B_x, \\
&\rho i\omega \delta u_y = -ikB\delta B_y, \\
  &i\omega \delta B_x = ikB\delta u_x + \eta_\mathrm{H} k^2\delta B_y, \\
  &i\omega \delta B_y = ikB\delta u_y - \eta_\mathrm{H} k^2\delta B_x.
\end{align}

The system of equations is linear and can be written in matrix form
\begin{equation}
\left( \begin{array}{cccc}  \rho i\omega & 0 & ikB & 0 \\   0 & \rho i\omega & 0 & ikB   \\  -ikB & 0 & i\omega & -\eta_\mathrm{H}k^2   \\ 0 & -ikB & \eta_\mathrm{H}k^2 & i\omega \end{array}\right) \left( \begin{array}{c} \delta u_x \\ \delta u_y \\ \delta B_x \\ \delta B_y \end{array} \right)  = 0.
\end{equation}

A non-trivial solution is possible if the matrix determinant is zero. We obtain
\begin{equation}
\omega^4 - \omega ^2 (2k^2c_\mathrm{A}^2 + \eta_\mathrm{H}^2k^4) + k^4 c_\mathrm{A}^4 =0,
\end{equation}
with $c_\mathrm{A}=\frac{B}{\sqrt{\rho}}$ the Alfv\'en speed.
This equation can also be written as
\begin{align}
  (\omega^2-k^2c_\mathrm{A}^2)^2 - (\eta_\mathrm{H}k^2\omega)^2 &= 0, ~\mathrm{then}\\
  (\omega^2 - \eta_\mathrm{H} k^2\omega - k^2 c_\mathrm{A}^2)(\omega^2 + \eta_\mathrm{H} k^2\omega - k^2 c_\mathrm{A}^2)&=0.
\end{align}
We only keep the positive solutions for $\omega$, which yields the following dispersion relation \citep{balbus2001}
\begin{equation}\label{disprel}
\omega = \pm \frac{\eta_\mathrm{H} k^2}{2} + \sqrt{\left(\frac{\eta_\mathrm{H} k^2}{2}\right)^2 + k^2c_\mathrm{A}^2}. 
\end{equation}
Contrarily to the ambipolar and the Ohmic diffusions, the Hall effect is purely dispersive. Any perturbation creates two waves whose frequencies $\omega$ are linked to their wavelength $\lambda = 2\pi /k$. They are called whistler waves. They propagate the magnetic field perturbations along the field lines at the whistler speed
\begin{equation}
  c_\mathrm{w} = \frac{\omega}{k} = \pm \frac{\eta_\mathrm{H} k}{2} + \sqrt{\left(\frac{\eta_\mathrm{H} k}{2}\right)^2 + c_\mathrm{A}^2}. \label{wspeed}
\end{equation}
One of these two waves is always faster than the Afv\'en speed, and can theoretically travel at an unbound speed for very high frequencies and very small wavelengths. In practice, it is physically limited by the ion cyclotron frequency \citep{Srinivasan}. From Equation (\ref{wspeed}), we can define two regimes. If $\frac{\eta_\mathrm{H} k}{2} \ll c_\mathrm{A}$ then the Hall effect is weak and $c_\mathrm{w} \approx c_\mathrm{A}$. In this case, the whistler waves behave as Alfv\'en waves. Otherwise, the Hall effect plays a significant role on the magnetic field evolution. This defines the Hall length $l_\mathrm{H} = \frac{\eta_\mathrm{H}}{c_\mathrm{A}}$ below which the Hall effect is dominant.
Note that by being purely dispersive, the Hall effect does not directly impact the fluid energy or momentum. It only modifies the topology of the magnetic field.

\section{Numerical methods - implementation of the Hall effect}\label{sect_implement}

We implemented the Hall effect in the Eulerian code {\ttfamily RAMSES} \citep{teyssier}. Since the Hall effect is dispersive, the introduction of new waves in the system modifies the Riemann problem that needs to be solved on each cell interface. For this reason we have adapted the method of \citet{Lesur2014} implemented in the PLUTO code \citep{2007ApJS..170..228M}. This method is similar to \citet{Toth2008} and is based on a direct modification of the Riemann solver. Therefore, we do not consider the Hall effect as a source term, as it is the case for the ambipolar and Ohmic diffusions in {\ttfamily RAMSES} \citep{masson_nimhd}.

\subsection{The constrained transport in {\ttfamily RAMSES}}

\begin{figure}
\begin{center}
\includegraphics[width=0.5\textwidth]{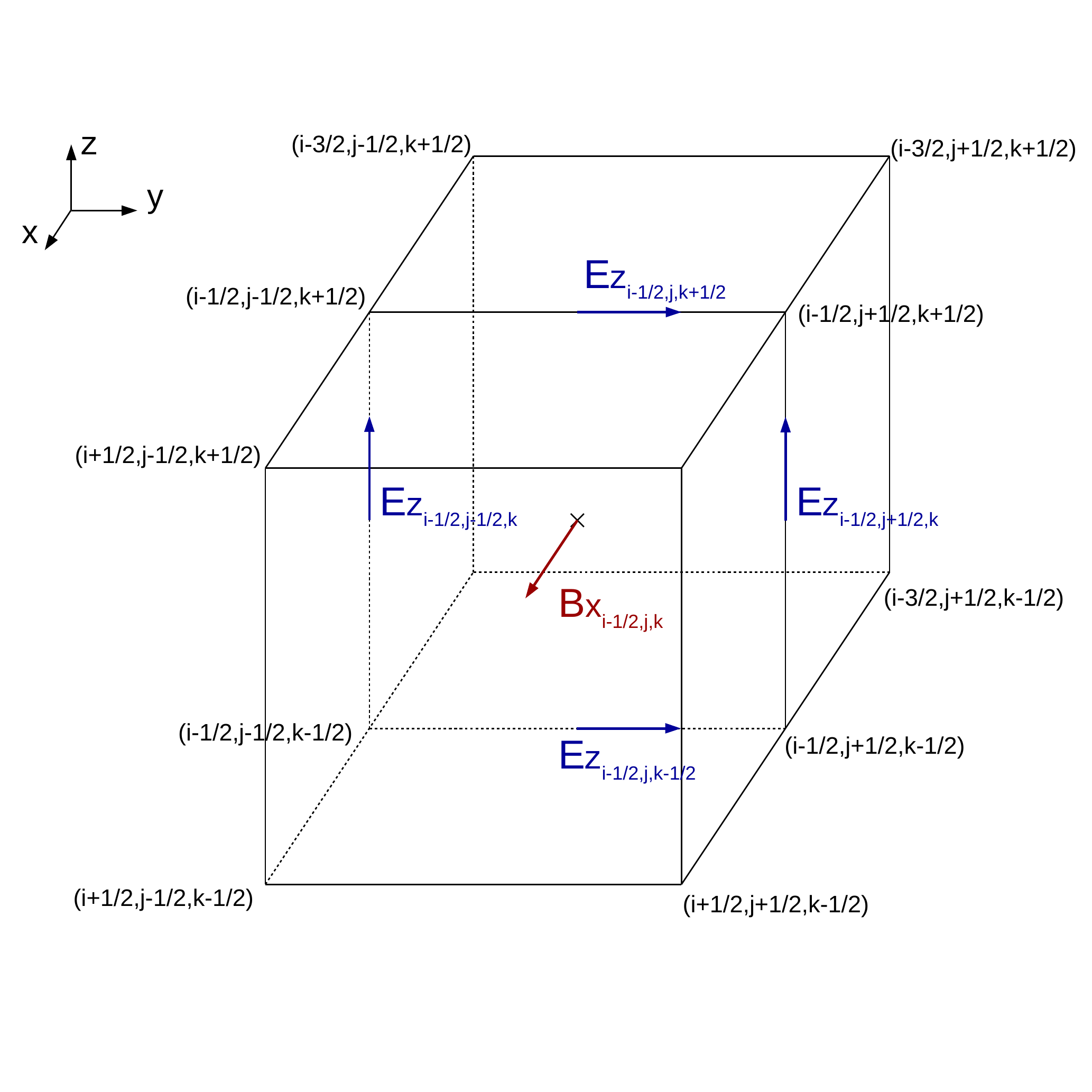}
\caption{Visualization of the constrained transport (CT) method for the calculation of the magnetic field $B_x$ at cell interface $(i-\frac{1}{2},j,k)$.}
\label{stokes}
\end{center}
\end{figure}

In {\ttfamily RAMSES}, the magnetic field is naturally defined at the center of each orthogonal face, meaning that for the cell centered on indexes $(i,j,k)$ (for the $x$, $y$ and $z$ directions), $B_x$, $B_y$, and $B_z$ are defined at $(i-\frac{1}{2},j,k)$, $(i,j-\frac{1}{2},k),$ and $(i,j,k-\frac{1}{2})$ respectively. The code uses the constrained transport (CT) \citep{1988ApJ...332..659E} to solve both the induction equation of ideal MHD \citep{teyssierMHD,fromang} and the ambipolar and Ohmic diffusions \citep{masson_nimhd}. This method exploits the Stokes theorem by using the value of the electromotive forces (EMF) on cell edges to compute the magnetic flux through cell interfaces. For instance, Figure \ref{stokes} shows which EMFs are used to evolve the $x-$component of the magnetic field at the cell interface $(i-1/2,j,k)$. The constrained transport update reads
\begin{align}
  \frac{B^{n+1}_{x,i-\frac{1}{2},j,k}-B^{n}_{x,i-\frac{1}{2},j,k}}{\Delta t} = &\frac{E^{n+\frac{1}{2}}_{z,i-\frac{1}{2},j+\frac{1}{2},k}-E^{n+\frac{1}{2}}_{z,i-\frac{1}{2},j-\frac{1}{2},k}}{\Delta y} \nonumber \\
                                                                               &- \frac{E^{n+\frac{1}{2}}_{y,i-\frac{1}{2},j,k+\frac{1}{2}}-E^{n+\frac{1}{2}}_{y,i-\frac{1}{2},j,k-\frac{1}{2}}}{\Delta z}.\label{inducfinite}
\end{align}
In {\ttfamily RAMSES}, since $\Delta x=\Delta y=\Delta z$, we use only $\Delta x$ in the remainder of the manuscript.
As {\ttfamily RAMSES} uses the MUSCL scheme \citep{1976cppa.conf...E1V}, a predictive step is first performed to compute the flow variables from time-step $n$ to $n+1/2$. These predictions are then used to obtain the fluxes (here the EMFs). In the case of the CT, the EMFs are located at cell edges, at the intersection of four different states. In {\ttfamily RAMSES}, a 2D-Riemann solver is used to deal with the discontinuity and compute the EMFs from these states. We explicitly detail these steps in the description of our implementation.

\subsection{Principle of the implementation}\label{principle}

The Hall induction equation can be rewritten 
\begin{equation}
  \frac{\partial \mathbf{B}}{\partial t} = - \nabla \times [(\mathbf{u}+ \mathbf{u}_\mathrm{H}) \times \mathbf{B}],
\end{equation}
where $\mathbf{u}_\mathrm{H} \equiv - \eta_\mathrm{H} \frac{\nabla \times \mathbf{B}}{||\mathbf{B}||}$ is the Hall speed (not to be confused with the whistler speed !), at which the Hall effect moves the field lines. Our goal is to replace the fluid velocity $\mathbf{u}$ by the "effective" velocity $\mathbf{u}+\mathbf{u}_\mathrm{H}$ in the 2D-Riemann solver that computes the EMFs on cell edges for the CT scheme.

The Hall effect is a dispersive term which induces the propagation of waves at every wavelength. The Shannon-Nyquist theorem \citep{shannon} states that whistler waves with a wavelength smaller than twice the local resolution cannot be described and propagated. This defines the minimum resolved whistler wavelength as $\lambda = 2\Delta x$, where $\Delta x$ is the size of the local cell. Following Equation \eqref{wspeed}, the associated whistler speed is
\begin{equation}\label{whistlernum}
  c_\mathrm{w} = \frac{\eta_\mathrm{H}\pi}{2\Delta x} + \sqrt{\left(\frac{\eta_\mathrm{H}\pi}{2\Delta x}\right)^2 + c_\mathrm{A}^2}.
\end{equation}
Since {\ttfamily RAMSES} is an explicit code in time, this wave speed needs to be accounted for in the computation of the characteristic speeds. $c_\mathrm{w}$ however scales as $1/\Delta x$, which consequently decreases the spatial order of the scheme by one. To get a consistent implementation, an at-least second-order scheme is hence necessary, which is the case here with the MUSCL scheme.

\subsection{The Hall speed}

Since the CT scheme uses the EMFs on cell edges, the Hall speed $\mathbf{u}_\mathrm{H}$ also needs to be computed on cell edges. We take the example of the edge located at $(i-\frac{1}{2},j-\frac{1}{2},k)$.
The first step is to determine the predicted states of the flow variables on cell faces and edges using a TVD slope. As part of the MUSCL scheme, the principle is to use a linear function whose slope is determined by considering neighboring cells. The TVD method \citep{1983JCoPh..49..357H} limits this slope to prevent the generation of new, unwanted, extrema in the system. For example, a well-known slope limiter is minmod. For the variable $U$ in the cell with index $i$, the minmod slope is 0 if $U_i$ is a local extremum, and the lowest magnitude between $(U_i-U_{i-1})$ and $(U_{i+1}-U_i)$ otherwise. For the magnetic field, the predicted states are computed using the induction equation without the Hall term, coupled with the Stokes theorem 
\begin{align}
  \frac{B^{n+\frac{1}{2}}_{x,i-\frac{1}{2},j,k}-B^{n}_{x,i-\frac{1}{2},j,k}}{\Delta t} =  &\frac{E^n_{z,i-\frac{1}{2},j+\frac{1}{2},k}-E^n_{z,i-\frac{1}{2},j-\frac{1}{2},k}}{\Delta x} \nonumber\\
                                                                - &\frac{E^n_{y,i-\frac{1}{2},j,k+\frac{1}{2}}-E^n_{y,i-\frac{1}{2},j,k-\frac{1}{2}}}{\Delta x}, \label{bstokes1}\\
  \frac{B^{n+\frac{1}{2}}_{y,i,j-\frac{1}{2},k}-B^{n}_{y,i,j-\frac{1}{2},k}}{\Delta t} =  &\frac{E^n_{x,i,j-\frac{1}{2},k+\frac{1}{2}}-E^n_{x,i,j-\frac{1}{2},k-\frac{1}{2}}}{\Delta x} \nonumber \\
                                                                - &\frac{E^n_{z,i+\frac{1}{2},j-\frac{1}{2},k}-E^n_{z,i-\frac{1}{2},j-\frac{1}{2},k}}{\Delta x}, \label{bstokes2}\\
  \frac{B^{n+\frac{1}{2}}_{z,i,j,k-\frac{1}{2}}-B^{n}_{z,i,j,k-\frac{1}{2}}}{\Delta t} =  &\frac{E^n_{y,i+\frac{1}{2},j,k-\frac{1}{2}}-E^n_{y,i-\frac{1}{2},j,k-\frac{1}{2}}}{\Delta x} \nonumber \\
                                                                - &\frac{E^n_{x,i,j+\frac{1}{2},k-\frac{1}{2}}-E^n_{x,i,j-\frac{1}{2},k-\frac{1}{2}}}{\Delta x}, \label{bstokes3}
\end{align}

where $\mathbf{E}^n=\mathbf{u}^n \times \mathbf{B}^n$ is the electric field calculated on cell edges using simple interpolation of $\mathbf{u}^n$ and $\mathbf{B}^n$ at these locations, for example,
\begin{align}
  \mathbf{u}^n_{i-\frac{1}{2},j-\frac{1}{2},k} &= \frac{1}{4} \left( \mathbf{u}^n_{i,j,k} + \mathbf{u}^n_{i-1,j,k} + \mathbf{u}^n_{i,j-1,k} +\mathbf{u}^n_{i-1,j-1,k} \right), \label{eqpred1}\\
  B^n_{x,i-\frac{1}{2},j-\frac{1}{2},k} & = \frac{1}{2} \left( B^n_{x,i-\frac{1}{2},j,k} + B^n_{x,i-\frac{1}{2},j-1,k} \right),\\
  B^n_{y,i-\frac{1}{2},j-\frac{1}{2},k} & = \frac{1}{2} \left( B^n_{y,i,j-\frac{1}{2},k} + B^n_{x,i-1,j-\frac{1}{2},k} \right).\label{eqpred2}
\end{align}

Only the perpendicular component of the magnetic field is calculated on each face. We reconstruct the other two by interpolating the state at the cell center using the TVD slope. For instance, obtaining $B_y$ on the face located at $(i-\frac{1}{2},j,k)$ yields
\begin{align}
  B^{n+\frac{1}{2}}_{y,i,j,k} &= \frac{1}{2}\left(B^{n+\frac{1}{2}}_{y,i,j-\frac{1}{2},k}+B^{n+\frac{1}{2}}_{y,i,j+\frac{1}{2},k} \right), \label{b1} \\
  B^{n+\frac{1}{2}}_{y,i-\frac{1}{2},j,k} &= B^{n+\frac{1}{2}}_{y,i,j,k} - \frac{\Delta_x(B_y)_{i,j,k}}{2},\label{b2}
\end{align}

where $\Delta_x(B_y)_{i,j,k}$ is the slope of $B_y$ in the $x-$direction for cell $(i,j,k)$.
As a result, two different magnetic fields are defined on each face, computed independently using the flow variables of the cells on each side of the interface. From here on, we have used the $LI$ and $RI$ exponents (left and right interfaces) to indicate the component coming from the lower index and higher index cells respectively, for example,

\begin{align}
  \mathbf{B}^{n+\frac{1}{2},LI}_{i-\frac{1}{2},j,k} &= \mathbf{B}^{n+\frac{1}{2}}_{i-1,j,k} + \frac{\Delta_x(\mathbf{B})_{i-1,j,k}}{2},\\
  \mathbf{B}^{n+\frac{1}{2},RI}_{i-\frac{1}{2},j,k} &= \mathbf{B}^{n+\frac{1}{2}}_{i  ,j,k} - \frac{\Delta_x(\mathbf{B})_{i  ,j,k}}{2},
\end{align}

with the same logic for $y-$ and $z-$ directions. We note that in each case, the component perpendicular to the face is naturally defined and does not need to be interpolated with the slope.
This discontinuity of $\mathbf{B}$ is a problem here since the Hall speed is proportional to the electric current $\mathbf{J}=\nabla \times \mathbf{B}$. The jumps at cell interfaces may be assimilated to perturbations with an infinitely small wavelength that would propagate at an infinite speed, which is of course unphysical. To tackle this problem, we averaged $\mathbf{B}$ and consider that $\mathbf{J}$ is constant on cell faces and edges \citep{Toth2008,Lesur2014}, at the price of a greater numerical diffusivity since we smoothed out discontinuities.
We note $\mathbf{B}^\mathrm{pred}$ the averaged magnetic field, therefore calculated as
\begin{equation}
  \mathbf{B}^\mathrm{pred}_{i-\frac{1}{2},j,k} = \frac{\mathbf{B}^{n+\frac{1}{2},LI}_{i-\frac{1}{2},j,k}+\mathbf{B}^{n+\frac{1}{2},RI}_{i-\frac{1}{2},j,k}}{2}. 
\end{equation}

\begin{figure}
\begin{center}
\includegraphics[width=0.5\textwidth]{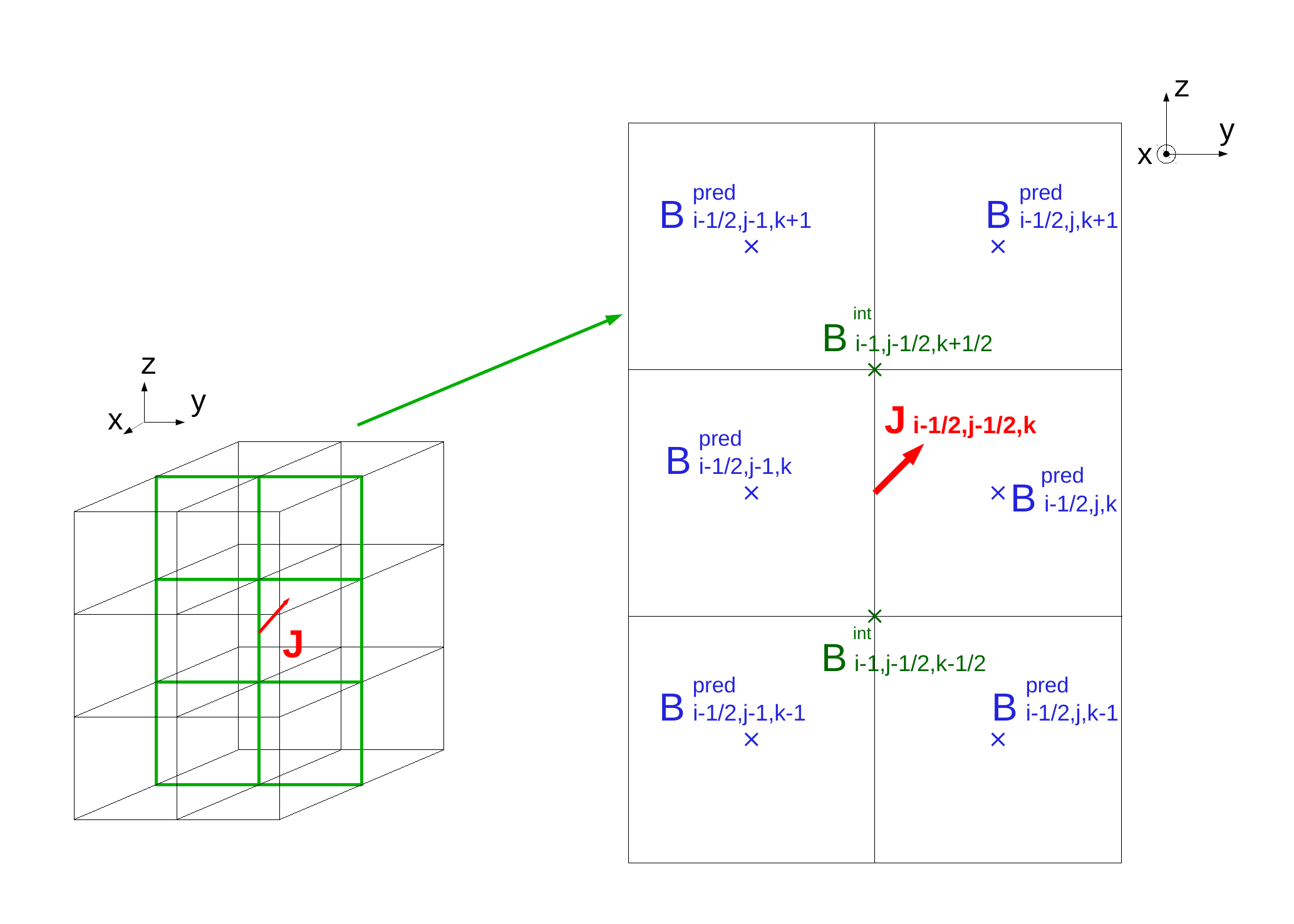}
\caption{Magnetic fields used to calculate $\mathbf{J}$ on edge $(i-\frac{1}{2},j-\frac{1}{2},k)$. The right picture represents the cell faces at $(i-\frac{1}{2})$, as highlighted on the 12 cells block $(i-1..i, j-1..j, k-1..k+1)$ on the left (the front right middle cell is cell $(i,j,k)$).}
\label{jcalc}
\end{center}
\end{figure}

Next we interpolated $\mathbf{J}$ on cell edges. For this, we also need the magnetic field at the two ends of the edges, that we interpolated from $\mathbf{B}^\mathrm{pred}$. On the edge located at $(i-\frac{1}{2},j-\frac{1}{2},k)$, this yields
\begin{align}
  \mathbf{B}^\mathrm{int}_{i-\frac{1}{2},j-\frac{1}{2},k+\frac{1}{2}} = \frac{\mathbf{B}^\mathrm{pred}_{i-\frac{1}{2},j-1,k}+\mathbf{B}^\mathrm{pred}_{i-\frac{1}{2},j-1,k+1}+\mathbf{B}^\mathrm{pred}_{i-\frac{1}{2},j,k+1}+\mathbf{B}^\mathrm{pred}_{i-\frac{1}{2},j,k}}{4}. \label{eqbint}\\
  \mathbf{B}^\mathrm{int}_{i-\frac{1}{2},j-\frac{1}{2},k-\frac{1}{2}} = \frac{\mathbf{B}^\mathrm{pred}_{i-\frac{1}{2},j-1,k-1}+\mathbf{B}^\mathrm{pred}_{i-\frac{1}{2},j-1,k}+\mathbf{B}^\mathrm{pred}_{i-\frac{1}{2},j,k}+\mathbf{B}^\mathrm{pred}_{i-\frac{1}{2},j,k-1}}{4}. \label{eqbint2}
\end{align}
Figure \ref{jcalc} depicts the magnetic fields used in the calculation of $\mathbf{J}$ on edge $(i-\frac{1}{2},j-\frac{1}{2},k)$. The faces represented are situated in $x=(i-\frac{1}{2})$ (see left picture).

We then computed $\mathbf{J}$ as $\nabla \times \mathbf{B}$

\begin{align}
  J_{x,i-\frac{1}{2},j-\frac{1}{2},k} = &\frac{B^\mathrm{pred}_{z,i-\frac{1}{2},j,k}-B^\mathrm{pred}_{z,i-\frac{1}{2},j-1,k}}{\Delta x} \nonumber \\
                                      - &\frac{B^\mathrm{int}_{y,i-\frac{1}{2},j-\frac{1}{2},k+\frac{1}{2}}-B^\mathrm{int}_{y,i-\frac{1}{2},j-\frac{1}{2},k-\frac{1}{2}}}{\Delta x}, \label{j1}\\
  J_{y,i-\frac{1}{2},j-\frac{1}{2},k} = &\frac{B^\mathrm{int}_{x,i-\frac{1}{2},j-\frac{1}{2},k+\frac{1}{2}}-B^\mathrm{int}_{x,i-\frac{1}{2},j-\frac{1}{2},k-\frac{1}{2}}}{\Delta x} \nonumber \\
                                      - &\frac{B^\mathrm{pred}_{z,i,j-\frac{1}{2},k}-B^\mathrm{pred}_{z,i-1,j-\frac{1}{2},k}}{\Delta x},\\
  J_{z,i-\frac{1}{2},j-\frac{1}{2},k} = &\frac{B^\mathrm{pred}_{y,i,j-\frac{1}{2},k}-B^\mathrm{pred}_{y,i-1,j-\frac{1}{2},k}}{\Delta x} \nonumber \\
                                      - &\frac{B^\mathrm{pred}_{x,i-\frac{1}{2},j,k}-B^\mathrm{pred}_{x,i-\frac{1}{2},j-1,k}}{\Delta x}.\label{j2}
\end{align}

The Hall resistivity $\eta_\mathrm{H}$ depends on the density $\rho$, the temperature $T$, the cosmic-ray ionization rate $\zeta$, and the magnetic field amplitude $B$. These quantities are interpolated on cell edges from the cell-centered predicted states. For each of these variables $A$, we computed
\begin{equation}
  \overline{A}_{i-\frac{1}{2},j-\frac{1}{2},k} = \frac{1}{4}\left(A^{n+\frac{1}{2}}_{i,j,k}+A^{n+\frac{1}{2}}_{i-1,j,k}+A^{n+\frac{1}{2}}_{i-1,j-1,k}+A^{n+\frac{1}{2}}_{i,j-1,k}\right).\label{average}
\end{equation}

Finally, the Hall speed is 
\begin{equation}\label{hallspeedscheme}
  \mathbf{u}_{\mathrm{H},i-\frac{1}{2},j-\frac{1}{2},k}=-\frac{\eta_\mathrm{H}(\overline{\rho},\overline{T},\overline{\zeta},\overline{B})}{\overline{B}}\mathbf{J},
\end{equation}
with every quantity calculated at $(i-\frac{1}{2},j-\frac{1}{2},k)$.

\subsection{Solving the Riemann problem} \label{section_riemann}

\begin{figure}
\begin{center}
\includegraphics[trim=3cm 1cm 3cm 2cm, clip, width=0.5\textwidth]{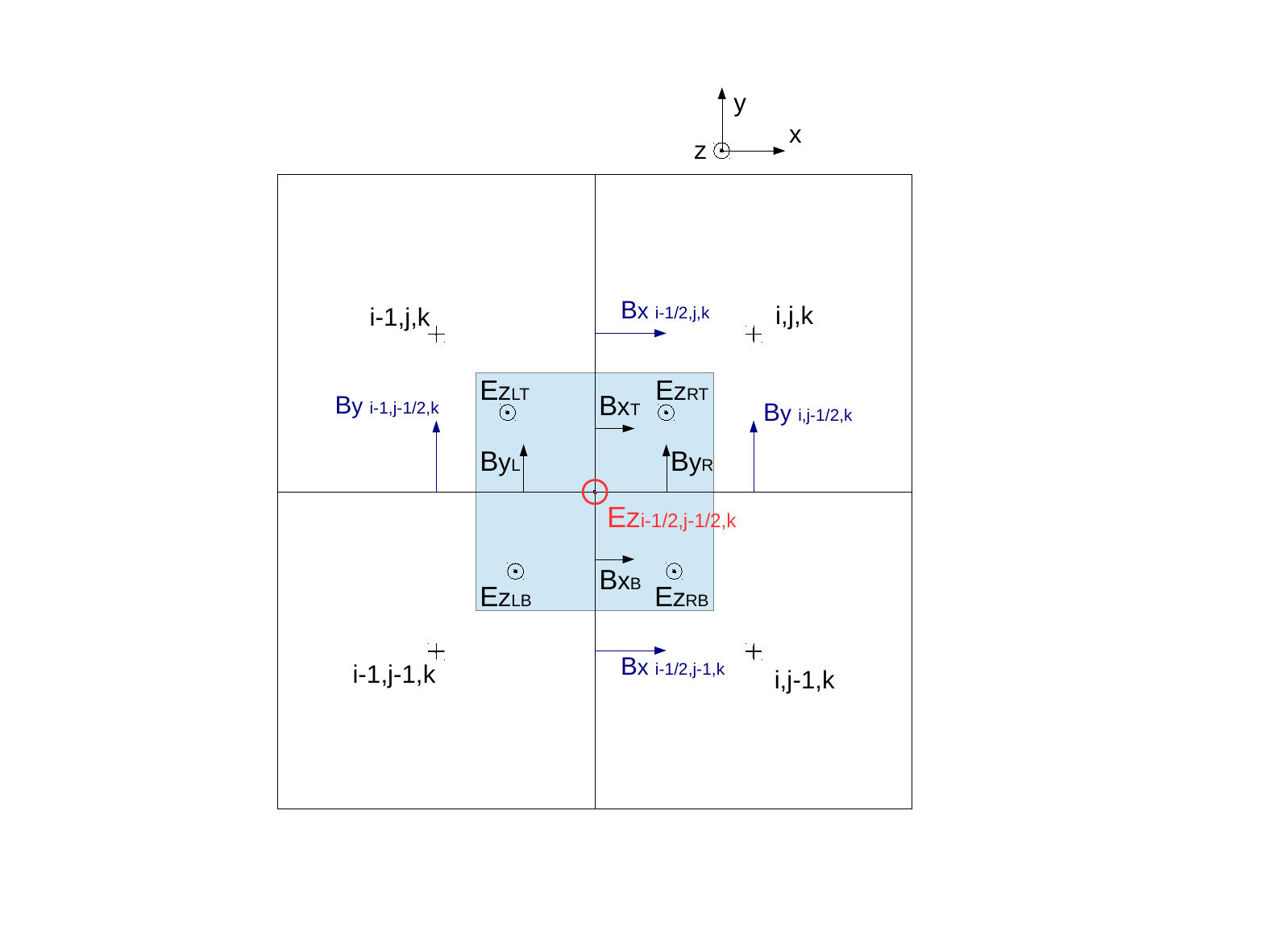}
\caption{Scheme of a 2D-Riemann problem in the z-direction. The blue area represents the states at the corner for each cell at $i-\frac{1}{2},j-\frac{1}{2},k$.}
\label{riemann2D}
\end{center}
\end{figure}

We then solved the 2D-Riemann problem to obtain the value of the total electromotive forces on cell edges. The notations are defined in Figure \ref{riemann2D}. The blue area represents the corner states of the four cells at $(i-\frac{1}{2},j-\frac{1}{2},k)$, as calculated in Equations \eqref{corn1} to \eqref{corn2}, and the red label is the emf computed in Equation \eqref{emfHLL}.

The flow variables were extrapolated on edges from the predicted states at cell centers using equations \eqref{testmass} to \eqref{testenergy} and the TVD slopes. Similarly to the magnetic field on cell faces, this process yields four different states at each corner
\begin{align}
  U^{n+\frac{1}{2},LB}_{i-\frac{1}{2},j-\frac{1}{2},k} &= U^{n+\frac{1}{2}}_{i-1,j-1,k} + \frac{\Delta_x(U)_{i-1,j-1,k}}{2} + \frac{\Delta_y(U)_{i-1,j-1,k}}{2}, \label{corn1}\\
  U^{n+\frac{1}{2},RB}_{i-\frac{1}{2},j-\frac{1}{2},k} &= U^{n+\frac{1}{2}}_{i,j-1,k}   - \frac{\Delta_x(U)_{i,j-1,k}}{2}   + \frac{\Delta_y(U)_{i,j-1,k}}{2},\\
  U^{n+\frac{1}{2},LT}_{i-\frac{1}{2},j-\frac{1}{2},k} &= U^{n+\frac{1}{2}}_{i-1,j,k}   + \frac{\Delta_x(U)_{i-1,j,k}}{2}   - \frac{\Delta_y(U)_{i-1,j,k}}{2},\\
  U^{n+\frac{1}{2},RT}_{i-\frac{1}{2},j-\frac{1}{2},k} &= U^{n+\frac{1}{2}}_{i,j,k}     - \frac{\Delta_x(U)_{i,j,k}}{2}     - \frac{\Delta_y(U)_{i,j,k}}{2},
\end{align}
for any flow variable $U$, and the magnetic field is averaged on the interfaces

\begin{align}
  B^{n+\frac{1}{2},B}_{x,i-\frac{1}{2},j-\frac{1}{2},k} &=  \frac{B^{n+\frac{1}{2},LB}_{x,i-\frac{1}{2},j-\frac{1}{2},k} +  B^{n+\frac{1}{2},RB}_{x,i-\frac{1}{2},j-\frac{1}{2},k}}{2} \nonumber\\
                                &=  \frac{ \left[B^{n+\frac{1}{2},LI}_{x,i-\frac{1}{2},j-1,k} + \frac{\Delta_y(B_x)_{i-1,j-1,k}}{2}\right] +\left[B^{n+\frac{1}{2},RI}_{x,i-\frac{1}{2},j-1,k} + \frac{\Delta_y(B_x)_{i,j-1,k}}{2}\right]}{2},
\end{align}
\begin{align}
  B^{n+\frac{1}{2},T}_{x,i-\frac{1}{2},j-\frac{1}{2},k} &=  \frac{B^{n+\frac{1}{2},LT}_{x,i-\frac{1}{2},j-\frac{1}{2},k} +  B^{n+\frac{1}{2},RT}_{x,i-\frac{1}{2},j-\frac{1}{2},k}}{2} \nonumber\\
                                &=  \frac{ \left[B^{n+\frac{1}{2},LI}_{x,i-\frac{1}{2},j,k} - \frac{\Delta_y(B_x)_{i-1,j,k}}{2}\right] +\left[B^{n+\frac{1}{2},RI}_{x,i-\frac{1}{2},j,k} - \frac{\Delta_y(B_x)_{i,j,k}}{2}\right]}{2},
\end{align}
\begin{align}
  B^{n+\frac{1}{2},L}_{y,i-\frac{1}{2},j-\frac{1}{2},k} &=  \frac{B^{n+\frac{1}{2},LB}_{y,i-\frac{1}{2},j-\frac{1}{2},k} +  B^{n+\frac{1}{2},LT}_{y,i-\frac{1}{2},j-\frac{1}{2},k}}{2} \nonumber\\
                                &=  \frac{ \left[B^{n+\frac{1}{2},LI}_{y,i-1,j-\frac{1}{2},k} + \frac{\Delta_x(B_y)_{i-1,j-1,k}}{2}\right] +\left[B^{n+\frac{1}{2},RI}_{y,i-1,j-\frac{1}{2},k} + \frac{\Delta_x(B_y)_{i-1,j,k}}{2}\right]}{2},
\end{align}
\begin{align}
  B^{n+\frac{1}{2},R}_{y,i-\frac{1}{2},j-\frac{1}{2},k} &=  \frac{B^{n+\frac{1}{2},RB}_{y,i-\frac{1}{2},j-\frac{1}{2},k} +  B^{n+\frac{1}{2},RT}_{y,i-\frac{1}{2},j-\frac{1}{2},k}}{2} \nonumber\\
                                                        &=  \frac{ \left[B^{n+\frac{1}{2},LI}_{y,i,j-\frac{1}{2},k} - \frac{\Delta_x(B_y)_{i,j-1,k}}{2}\right] +\left[B^{n+\frac{1}{2},RI}_{y,i,j-\frac{1}{2},k} - \frac{\Delta_x(B_y)_{i,j,k}}{2}\right]}{2}.\label{corn15}
\end{align}

We write these quantities $B_x^B$, $B_x^T$, $B_y^L$, and $B_y^R$. 
The EMFs were then computed at each corner-state, with the addition of the Hall speed \eqref{hallspeedscheme} (that is the same for the four states), so that $\mathbf{E}=(\mathbf{u}+\mathbf{u}_\mathrm{H})\times \mathbf{B}$. On the $(i-\frac{1}{2},j-\frac{1}{2},k)$ edge, we needed the $z-$component
\begin{align}
  E^{n+\frac{1}{2},LB}_z = (u^{n+\frac{1}{2},LB}_x + u_{\mathrm{H},x}) B^{L}_y - (u^{n+\frac{1}{2},LB}_y + u_{\mathrm{H},y}) B^{B}_x,\label{corn152}\\
  E^{n+\frac{1}{2},LT}_z = (u^{n+\frac{1}{2},LT}_x + u_{\mathrm{H},x}) B^{L}_y - (u^{n+\frac{1}{2},LT}_y + u_{\mathrm{H},y}) B^{T}_x,\\
  E^{n+\frac{1}{2},RB}_z = (u^{n+\frac{1}{2},RB}_x + u_{\mathrm{H},x}) B^{R}_y - (u^{n+\frac{1}{2},RB}_y + u_{\mathrm{H},y}) B^{B}_x,\\
  E^{n+\frac{1}{2},RT}_z = (u^{n+\frac{1}{2},RT}_x + u_{\mathrm{H},x}) B^{R}_y - (u^{n+\frac{1}{2},RT}_y + u_{\mathrm{H},y}) B^{T}_x. \label{corn2}
\end{align}

The resulting EMF on the edge was obtained by solving the Riemann problem at this discontinuity of four different states.
The Hall effect is dispersive and introduces two waves into the system, for a total of ten
\begin{itemize}
  \item[-] Two Alfv\`en waves, that propagate the perturbations of the magnetic field, $\lambda= u \pm c_\mathrm{A}$,
  \item[-] Two fast magneto-sonic waves, representing the correlation between $\mathbf{B}$ and $\rho$ by the coupling between Lorentz force and thermal pressure,\\
    $\lambda = u \pm \sqrt{\frac{1}{2}(c_\mathrm{s}^2+c_\mathrm{A}^2) + \frac{1}{2}\sqrt{(c_\mathrm{s}^2+c_\mathrm{A}^2)-4c_\mathrm{s}^2 c_{\mathrm{A},x}^2}}$ \\
     (with $c_{\mathrm{A,x}}=B_x/\sqrt{\rho}$ the component of the Alfv\`en speed in the direction of propagation and $c_\mathrm{s}=\sqrt{\gamma P/\rho}$ the sound speed),
  \item[-] Two slow magneto-sonic waves, representing the anti-correlation between $\mathbf{B}$ and $\rho$ by the coupling between Lorentz force and thermal pressure, \\
    $\lambda = u \pm \sqrt{\frac{1}{2}(c_\mathrm{s}^2+c_\mathrm{A}^2) - \frac{1}{2}\sqrt{(c_\mathrm{s}^2+c_\mathrm{A}^2)-4c_\mathrm{s}^2 c_{\mathrm{A},x}^2}}$,
  \item[-] One contact discontinuity wave $\lambda=u$,
  \item[-] Two dispersive whistler waves.
\end{itemize}

Given the high number of waves, we used approximate Riemann solvers to compute the flux at cell faces and the EMF on cell edges. \citet{Toth2008} demonstrated that the Lax-Friedrich Riemann solver is inconsistent with the Hall effect, because the whistler speed scales as $c_\mathrm{w}\sim \Delta x^{-1}$. We chose to use the HLL solver \citep{HLL1983}, which considers only the lowest and highest (algebraic) wave speeds in each direction, and a constant state is assumed between these waves. It is a satisfying compromise between accuracy and numerical complexity, compared, for instance, to the HLLD solver \citep{2005JCoPh.208..315M} which takes five wave speeds into account.
In the case of a 2D-Riemann problem, the HLL solution for the EMF is \citep{2004JCoPh.195...17L} 
\begin{align}
  & E^{n+\frac{1}{2}}_{z} = \nonumber \\
  & \frac{S_LS_BE^{n+\frac{1}{2},LB}_{z}-S_LS_TE^{n+\frac{1}{2},LT}_{z}-S_RS_BE^{n+\frac{1}{2},RB}_{z}+S_RS_TE^{n+\frac{1}{2},RT}_{z}}{(S_R-S_L)(S_T-S_B)} \nonumber\\
                                &-\frac{S_TS_B}{S_T-S_B}(B^T_{x} - B^B_{x})+\frac{S_RS_L}{S_R-S_L}(B^R_{y} - B^L_{y})\label{emfHLL},
\end{align}

\noindent where $S_L$, $S_R$, $S_B$, and $S_T$ are the lowest and highest wave speeds in the $x$ and $y-$directions. 
Once this work has been done for every edge, the magnetic field is updated with the constrained transport method, using equation \eqref{inducfinite}.

\subsection{The speed of whistler waves}

In order to correctly propagate the perturbations caused by the Hall effect, the whistler wave speed \eqref{whistlernum} needs to be taken into account in the Courant Friedrichs Lewy (CFL) condition \citep{1928CFL}. This condition ensures that information does not travel further than one cell in one time-step, otherwise it would be lost since information exchange happens between adjacent cells.
At high resolution, we have roughly $c_\mathrm{w} \approx \frac{\eta_\mathrm{H}\pi}{\Delta x}$, and the corresponding time-step
\begin{equation}
  \Delta t_\mathrm{Hall} \approx \frac{\Delta x}{c_\mathrm{w}} \propto \Delta x^2.
\end{equation}
The scaling of the time-step in $\Delta x^2$ induces a fast decrease of the time-step as resolution increases, which is a limitation of current numerical simulation codes. Techniques that allow a faster integration such as Super Time-Stepping cannot be employed for the Hall effect because a larger time-step would cut off the high-frequency whistler waves.

At high resolutions, the fast whistler waves have the fastest characteristic speeds of the Riemann problem. We then have $S_L=u_x-c_\mathrm{w}$, $S_R=u_x+c_\mathrm{w}$, $S_B=u_y-c_\mathrm{w}$ and $S_T=u_y+c_\mathrm{w}$. This impacts the 2D-Riemann problem (as described in the implementation (Section \ref{section_riemann})), as well as the 1D-Riemann problems that are solved at cell interfaces during the prediction step of the MUSCL scheme.

\subsection{Summary}

The different steps of the Hall resolution are, in order
\begin{itemize}
  \item[-] Compute the predicted states of the flow variables (Eq. \eqref{bstokes1} - \eqref{eqpred2}),
  \item[-] Interpolate $\mathbf{B}$ on cell faces and corners with Equations \eqref{b1}-\eqref{eqbint2},
  \item[-] Calculate the electric current $\mathbf{J}$ on edges with Equations \eqref{j1}-\eqref{j2},
  \item[-] Calculate the Hall speed on edges with the averaged flow variables and the interpolated $\mathbf{J}$ (Eqs. \eqref{average} - \eqref{hallspeedscheme}),
  \item[-] Extrapolate the flow variables on cell corners (Eqs. \eqref{corn1}-\eqref{corn15}),
  \item[-] Use the effective speed $\mathbf{u}+\mathbf{u}_\mathrm{H}$ instead of $\mathbf{u}$ to compute the EMFs at cell corners (Eqs. \eqref{corn152}-\eqref{corn2})
  \item[-] Solve the 2D-Riemann problem by considering the whistler waves in the calculation of $S_L$, $S_R$, $S_B$, and $S_T$, to get the electric fields $E^{n+1/2}$ with Equation \eqref{emfHLL},
  \item[-] Update the flow variables with the constrained transport scheme \eqref{inducfinite}.
\end{itemize}

\subsection{Comparison to other implementation}

Our implementation of the Hall effect in {\ttfamily RAMSES} followed the work of \citet{Lesur2014}, who included it in the Eulerian code PLUTO \citep{2007ApJS..170..228M}. While both codes use the CT algorithm, \citet{Lesur2014} use a 1D-Riemann solver to compute the EMF while we have included the Hall effect into 2D-Riemann solver, which, to the best of our knowledge, has never been done before. Both codes use a second-order scheme (Runge-Kutta for PLUTO), and share the same convergence properties for the Hall effect.
The approach is also similar to the implementation of \citet{Toth2008} for the BASTRUS code \citep{Toth:2006:PET:1217520.1217543}, though they implemented the Hall effect using an explicit time integration within an implicit code. Since implicit schemes are unconditionnally stables, they are not strictly bound to the whistler speed to choose their time-step. This emancipation results in a faster code because larger time-steps are allowed. Furthermore, without the constraint of the whistler speed, the scheme order is increased by one. The drawback of the BASTRUS code is a less precise scheme at small scales because the shortest whistler waves cannot be correctly propagated.

The Hall effect has been implemented in the Eulerian code ZeusTW \citep{StoneNorman,Krasnopolsky} according to \citet{2002ApJ...570..314S}. The term $(\nabla \times \mathbf{B}) \times\mathbf{B}$ is evolved at $t^{n+1/2}$ using a finite difference scheme to evaluate the curls in an explicit way. The resulting Hall-EMF is then added to the ideal EMF $\mathbf{u}\times \mathbf{B}$ into a modified constrained transport method. The tests in \citet{2002ApJ...570..314S} show that the dispersion relation of the Hall effect is verified, but we have not found any information about the accuracy of the scheme.

Due to their nature, Lagrangean codes cannot use the constrained transport method. In the SPH code PHANTOM \citep{2014MNRAS.444.1104W,2017ascl.soft09002P} as well as in \citet{Tsukamoto2015} and \citet{2017PASJ...69...95T}, the three non-ideal MHD terms were implemented as source terms. Testing them with the ambipolar diffusion shows second order convergence, but no information is given for the Hall effect, nor  the propagation of whistler waves at various resolution. They do, however, recover the theoretical dispersion relation, and perform the shock-test described in section \ref{shocktest} with results fitting the expected profile within 3\% \citep{Wurster2015}.

\section{Tests} \label{sect_tests}

In the following section, we describe how we tested our implementation of the Hall effect, especially the propagation of the whistler waves.
Except in the shock test, $\mathbf{E}_\mathrm{NIMHD}$ is always reduced to the Hall effect.

\subsection{Propagating whistler waves}

\subsubsection{Dispersion relation test} \label{reldispsection}

We tested the propagation of whistler waves with our scheme by following the setup described in \citet{Kunzlesur2013}. First, we aimed to retrieve the dispersion Relation \eqref{disprel}.
In a cubic box of length 1~cm with uniform density and pressure, we set a whistler wave with a wave-number $k$, and study its propagation in the $x$-direction and its oscillations. The initial MHD quantities are
\begin{align*}
  \rho &= 1~ \mathrm{g~cm}^{-3},                        &~~~ B_x &= 0.1 ~\mathrm{G},\\
  P &= 1.5 \times 10^{-5}~\mathrm{dyne~cm}^{-2},        &~~~ B_y &= 10^{-3} \cos(2\pi kx) ~\mathrm{G},\\
  \gamma &= \frac{5}{3},                                & ~~~B_z &= 0 ~ \mathrm{G},\\
  \mathbf{u} &= 0 ~ \mathrm{cm~s}^{-1}.                 & ~~~&~~~
\end{align*}
We imposed periodic boundary conditions, which requires $k$ to be an integer to ensure continuity. To retrieve the dispersion Relation \eqref{disprel}, we vary $k$ and $\eta_\mathrm{H}$, typically between $5$ and $20$ for $k$ and $5\times10^{-4}$ to $0.1$ cm$^2$ s$^{-1}$ for $\eta_\mathrm{H}$. We used a uniform resolution of $\Delta x = 1/128$ cm and the moncen slope limiter. The question of resolution and slope limiters are discussed in the next section.

\begin{figure}
\begin{center}
\includegraphics[trim= 2cm 1cm 2cm 2cm, width=0.49\textwidth]{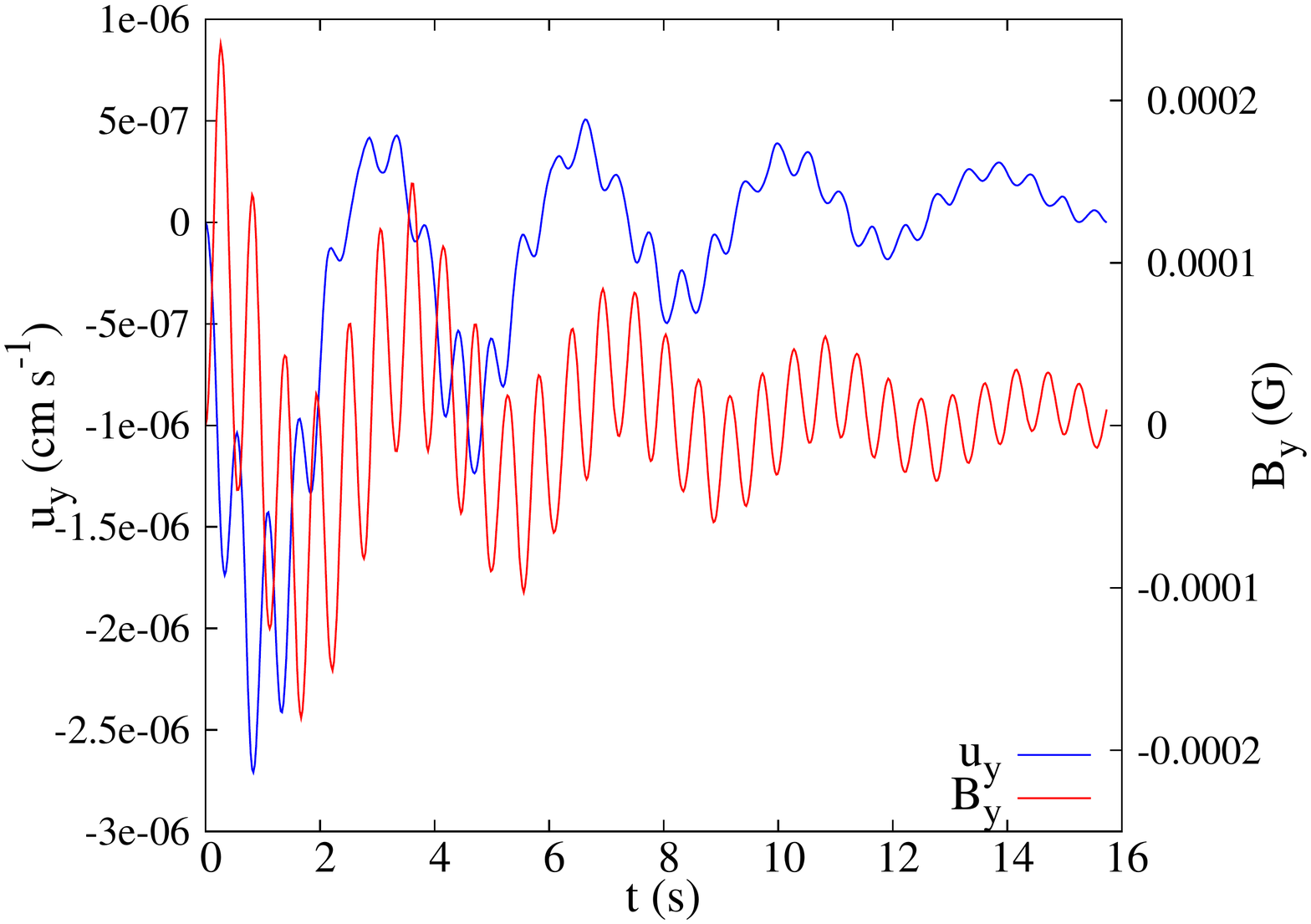}
\caption{Typical evolution of $u_y$ (in blue) and $B_y$ (in red) through time for a random cell in the box (here $x=48.5/128, ~y=0.5/128, ~z=63.5/128$). $k=7$ cm$^{-1}$, $\eta_\mathrm{H}=5\times 10^{-3}$ cm$^2$ s$^{-1}$.}
\label{vversust}
\end{center}
\end{figure}

\begin{figure}
\begin{center}
  \includegraphics[trim= 1cm 1cm 1cm 1cm, width=0.49\textwidth]{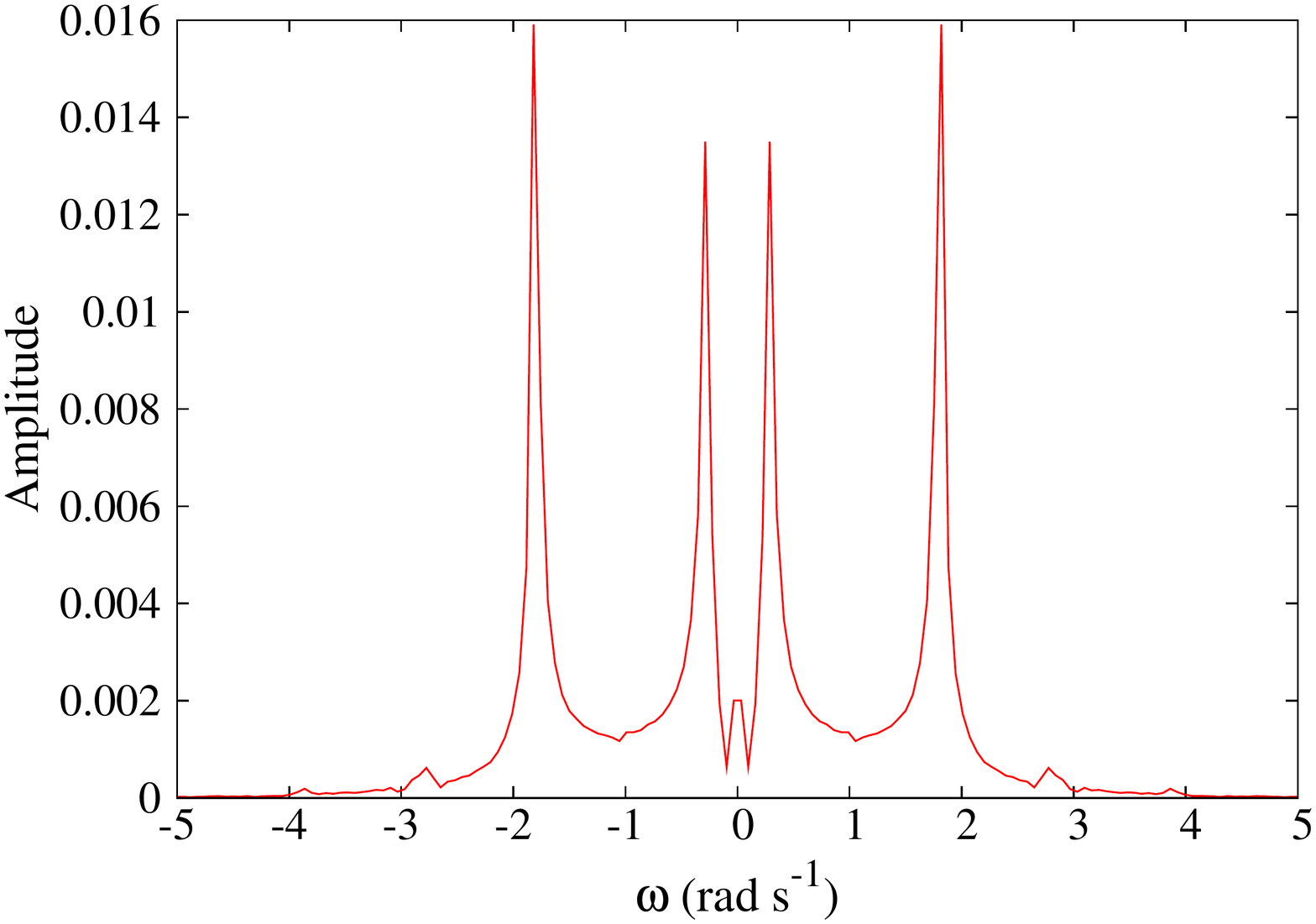}
  \caption{Fourier transform of $B_y(t)$ for a random cell in the box (same as figure \ref{vversust}).}
\label{fourier}
\end{center}
\end{figure}

\begin{figure}
\begin{center}
\includegraphics[trim= 1cm 1cm 1cm 1cm, width=0.49\textwidth]{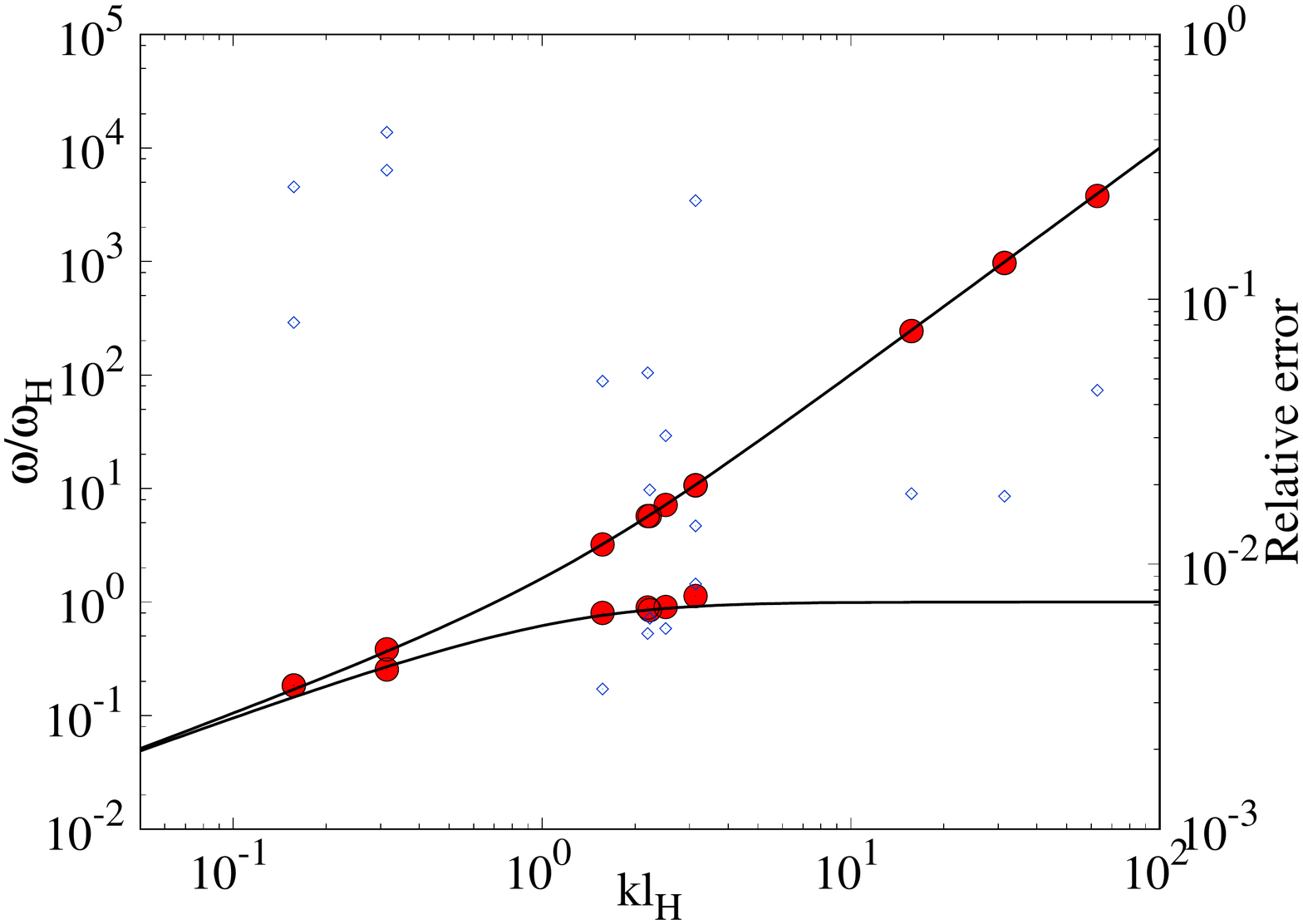}
\caption{Correspondence between the theoretical dispersion relation of the whistler waves (solid line), and the main oscillation frequencies of the magnetic field (circles). The blue square points represent the relative error.}
\label{reldisp1}
\end{center}
\end{figure}

Figure \ref{vversust} shows the time evolution of $u_y$ and $B_y$ in the case $k=7$ and $\eta_\mathrm{H}=5\times 10^{-3}$ cm$^2$ s$^{-1}$, for a random cell in the computational domain (here $x=y=32.5/128,~z=63.5/128$). 
As expected, two whistler waves propagate with different frequencies. The fluid is set into motion at the same frequencies as the magnetic field. Figure \ref{fourier} shows the Fourier transform of the temporal evolution of the magnetic field in the $y$ direction for the same cell as Fig. \ref{vversust}. It contains two peaks corresponding to the frequencies at which the waves propagate. These peaks are visible around $\omega \approx 2$ s$^{-1}$ and $\omega \approx 0.3$ s$^{-1}$.
By varying the Hall resistivity $\eta_\mathrm{H}$ and the wave number $k$, we obtain a set of frequencies. Figure \ref{reldisp1} shows the frequencies, normalized by the Hall frequency $\omega_\mathrm{H} = \frac{c_\mathrm{A}^2}{\eta_\mathrm{H}}$, plotted as a function of $kl_\mathrm{H}$. The agreement between our results and the theoretical dispersion relation is very good (within 5\% for most points, see blue points in Figure \ref{reldisp1}). However, we could not extract good approximations of the low frequency waves at high $kl_\mathrm{H}$ because of the fast dissipation of the signal (see Sect. 4.1.2). For a similar reason, the eigenfrequencies of $kl_\mathrm{H} < 1$ waves are less precise because of their slow oscillations.

\subsubsection{Truncation errors and order of the scheme}

Numerical schemes are based on the approximation of Taylor series, which creates truncation errors. In the context of wave propagations, they take the form of an artificial dissipation of the signal. For a well-designed numerical scheme, this error should decrease when the resolution increase. The damping of the signal clearly appears in Fig. \ref{vversust} with the decrease of the wave amplitudes.

\begin{figure}
\begin{center}
  \includegraphics[trim= 1cm 1cm 1cm 1cm, width=0.49\textwidth]{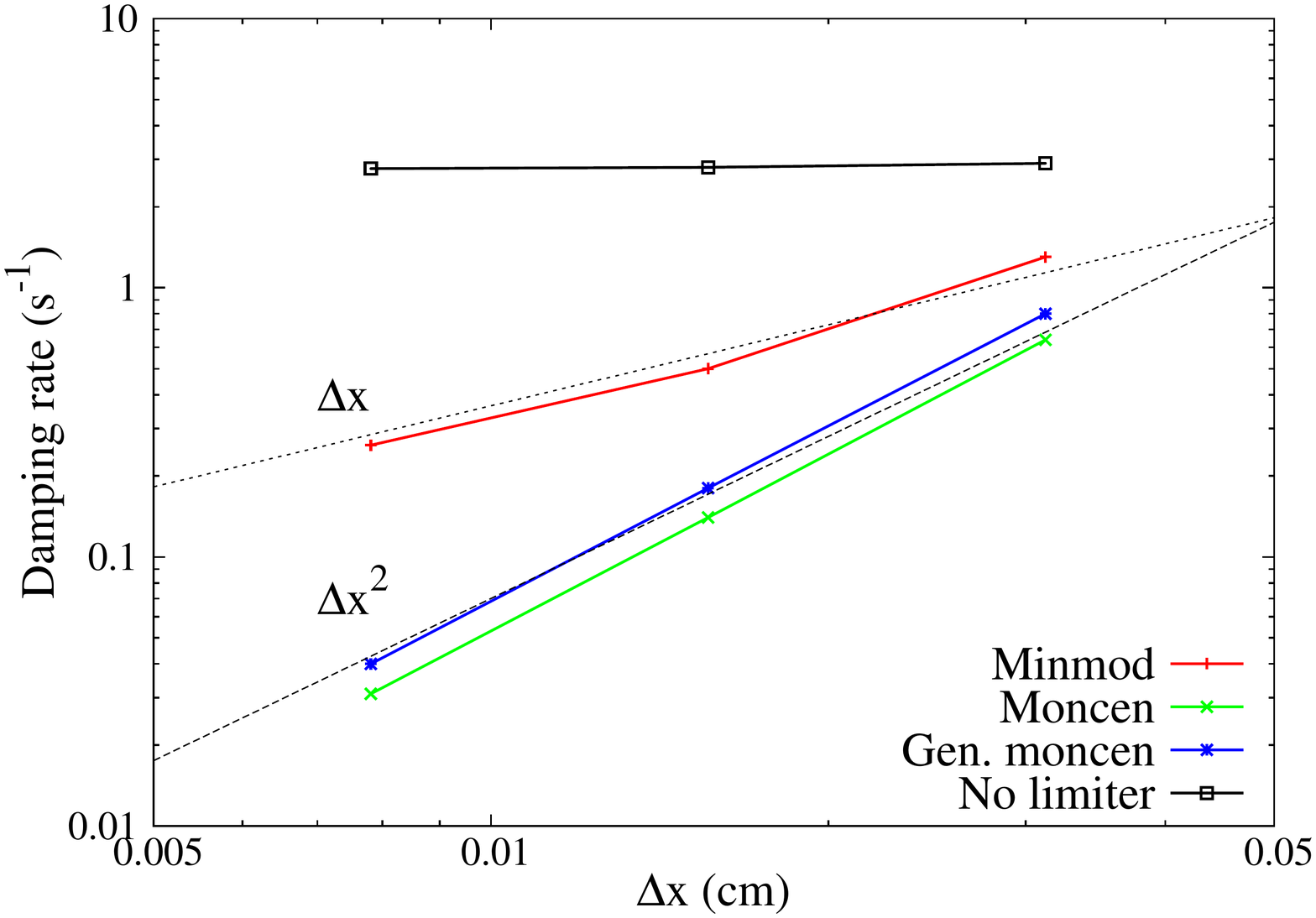}
\caption{Numerical damping rates as a function of resolution for different slope limiters, and $\Delta x$ and $\Delta x^2$ scalings for comparison.}
\label{Orderfigure}
\end{center}
\end{figure}

The transverse magnetic field can then be estimated as
\begin{equation}
B_y (t) = f(t) e^{-at},
\end{equation}
where $f(t)$ is the periodic contribution and $e^{-at}$ the damping, with $a$ the damping coefficient in s$^{-1}$. Ideally, this coefficient needs to be as small as possible. To achieve this, we wanted to make sure that our scheme is at the highest possible order in space, meaning, at least second-order in our implementation. We were able to control the convergence of the calculation by changing the grid resolution, as well as the slope limiter as suggested in \citet{Toth2008} and \citet{Lesur2014}.
We use the same setup as previously, with $k=5$ cm$^{-1}$, $\eta_\mathrm{H}=0.005$ cm$^2$ s$^{-1}$. We also used the slope limitersminmod, moncen, and generalized moncen ($\beta = 1.5$), for resolutions of $\Delta x= \frac{1}{32},~\frac{1}{64}~\mathrm{and}~\frac{1}{128}$ cm. The resulting damping rates $a$ are plotted in Figure \ref{Orderfigure}. Similarly to \citet{Toth2008} and \citet{Lesur2014}, we find that the minmod limiter is only first-order, while the moncen and generalized moncen limiters show second-order accuracy.

A second-order scheme like MUSCL, combined with the appropriate slope limiter, appears to be necessary to treat the Hall effect. As already mentioned in Section \ref{principle}, the characteristic whistler speed scales as $1/\Delta x$, which lowers the order of the scheme by one. The test results show that without a TVD slope, the numerical dissipation does not decrease with the resolution, and the scheme is inconsistent. 

\begin{figure}
\begin{center}
  \includegraphics[trim= 1cm 1cm 1cm 1cm, width=0.49\textwidth]{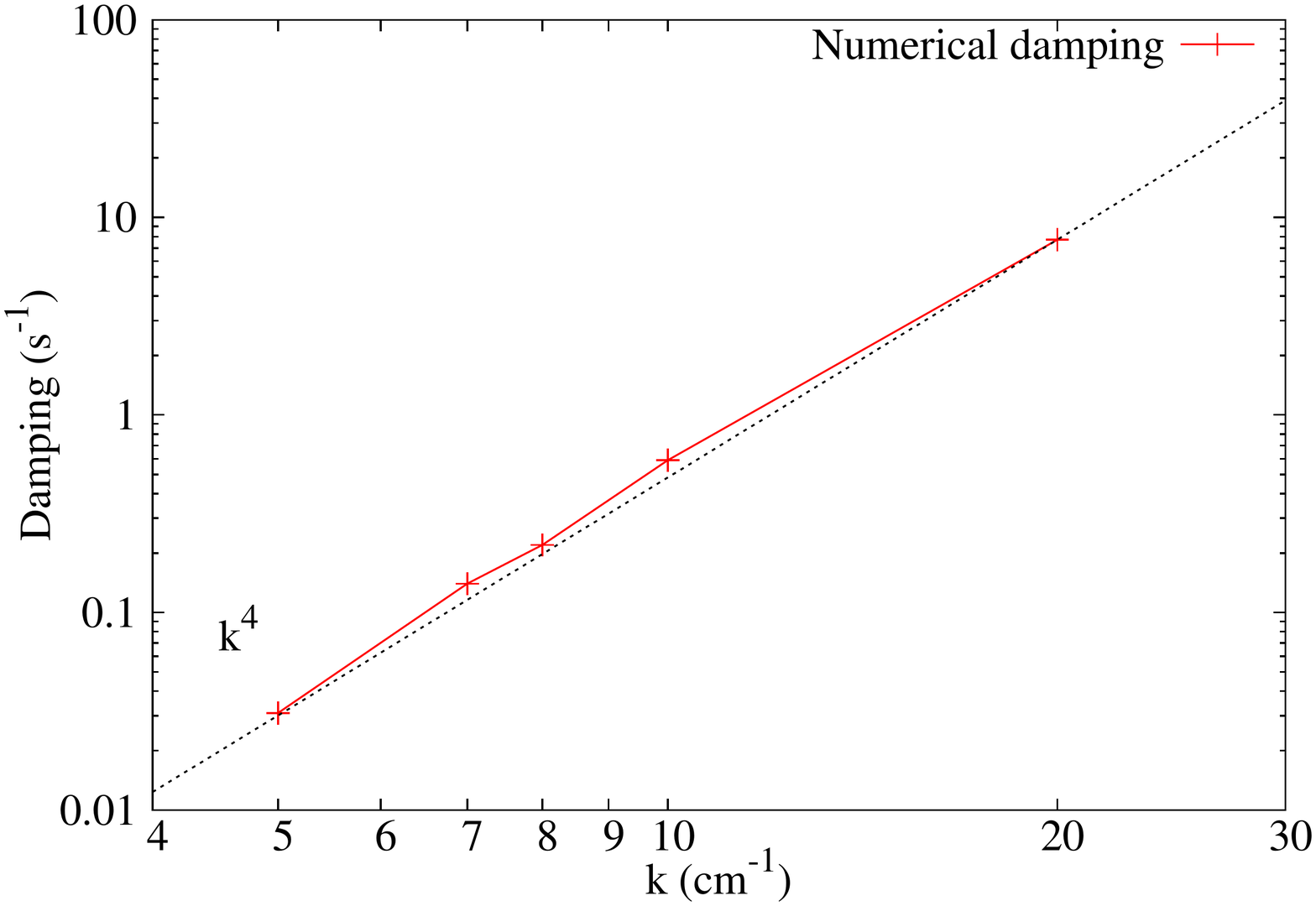}
\caption{Numerical damping rates versus wave-number $k$. Solid line: numerical results, dashed line: $k^4$ scaling.}
\label{damping}
\end{center}
\end{figure}

\citet{Toth2008} show that for a second-order slope limiter, the numerical dissipation scales as $\frac{\partial^4U}{\partial x^4}$. It means that at a given number of points per $l_\mathrm{H}$, the damping rate increases as $k^4$. We evaluated the damping rate for $\eta_\mathrm{H}=0.005$ cm$^2$ s$^{-1}$ at a fixed resolution of $\Delta x = \frac{1}{128}$ cm and $k=5,~7,~8,~10,~20$ cm$^{-1}$. We used the moncen slope limiter. Figure \ref{damping} shows the evolution of the numerical damping rate as a function of the wave-number $k$. We recovered the theoretical scaling in $k^4$, which indicates that smaller wavelengths are substantially more affected by the numerical dissipation than the larger ones.

\subsubsection{Grid refinement}
Here we investigate the behavior of whistler waves on non-regular grids, using the same plane-parallel setup of the previous tests, while refining the grid on specific locations. We once more took $k=5$~cm$^{-1}$ and $\eta_\mathrm{H}=0.005$ cm$^2$ s$^{-1}$ as the reference case, with the moncen slope limiter. We typically used two or three different levels of refinement, with the maximum resolution centered on $x=0$ and gradually decreasing on each side. The simulation box extends between $x=-0.5$~cm and $x=+0.5$~cm. A refinement level $\ell$ corresponds to a resolution $\Delta x = 1/2^\ell$ cm. For the two-levels tests, the levels of resolution are $\ell=\ell_\mathrm{min}+1$ for $-0.25 < x < 0.25$, and $\ell=\ell_\mathrm{min}$ everywhere else. For three-levels tests, we impose $\ell=\ell_\mathrm{min}+2$ between $-0.125 < x < 0.125$, $\ell=\ell_\mathrm{min}+1$ in $-0.25 < x < 0.25$, and $\ell=\ell_\mathrm{min}$ in the rest of the box. We chose $\ell_\mathrm{min} = 5,6$ for the two-levels simulations and $\ell_\mathrm{min}=4,5,6$ for the three-levels. The whistler wave propagates along the $x$-axis.

\begin{figure}
\begin{center}
\includegraphics[trim= 1cm 1cm 1cm 1cm, width=0.49\textwidth]{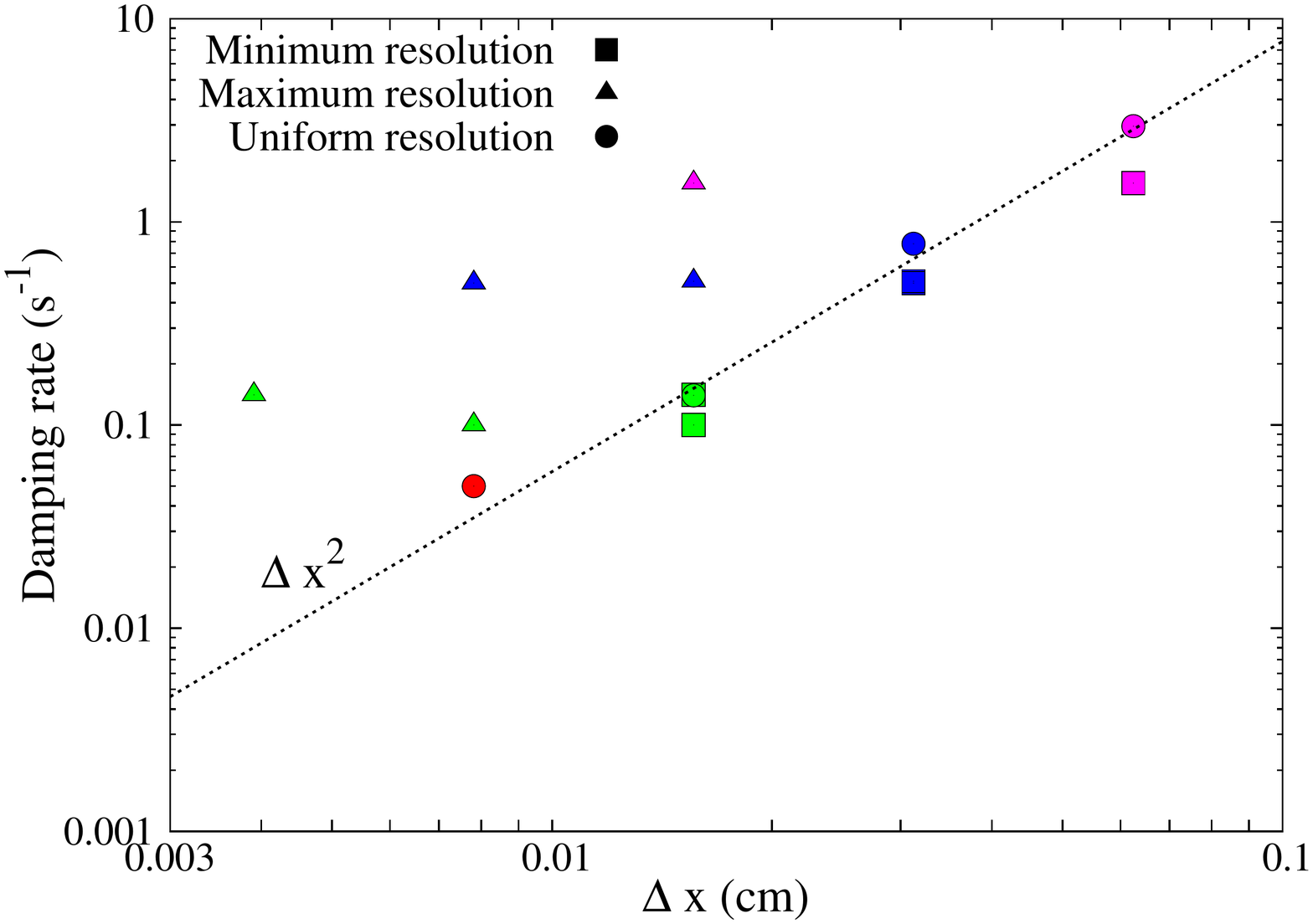}
  \caption{Damping coefficient of the whistler wave as a function of the maximum (triangles) and minimum (squares) resolution of the grid. Colors indicate the minimum resolution: purple for $\ell_\mathrm{min}=4$, blue for $\ell_\mathrm{min}=5$, green for $\ell_\mathrm{min}=6$, and red for $\ell_\mathrm{min}=7$. The circle points represent the regular grids, and the dashed line the least-squares fit of these points ($\propto \Delta x^{2.11}$).}
\label{dampamr}
\end{center}
\end{figure}

Figure \ref{dampamr} shows the damping rates of the whistler waves as a function of the maximum (triangles) and minimum (squares) resolution for the five runs described above, and for regular grids (circles) with minimum refinement levels from four to seven (colors). The damping in regular grids follows the same $\Delta x^2$ scaling as in the previous test. However, non-regular grids produce a higher damping rate that does not always scale down with the maximum resolution. Instead, the damping rate seems to be only affected by the minimum resolution. While the AMR is an efficient tool to capture small scale phenomena at a relatively low cost, it does not seem to be adapted for the propagation of whistler waves, whose dissipation depends only on the coarser resolution they cross.

\subsection{Shock test}\label{shocktest}

We then tested the Hall effect in a more physical situation. We considered a unidimensional isothermal shock in a Hall-dominated regime in presence of the Ohmic and the ambipolar diffusions, following the setup of \citet{Falle2003} and \citet{Wurster2015}.
The sound speed and $x$-magnetic field are uniform, with $c_\mathrm{s} = \sqrt{\frac{P}{\rho}} = 0.1$ cm s$^{-1}$, $B_x=1$ G, and we used the resistivities $\eta_\Omega = 10^{-9}$, $\eta_\mathrm{H} = -2 \times 10^{-2} B$ and $\eta_\mathrm{AD} = 3.5 \times 10^{-3} \frac{B^2}{\rho}$ cm$^2$ s$^{-1}$. We considered a cubic box with a discontinuity in the $x$-direction, and characterized by the left and right states presented in Table \ref{Shocktable}
\begin{table}
  \caption{Initial and boundary flow variables for the shock test}
  \label{Shocktable}
  \centering
\begin{tabular}{lrr}
\hline\hline
  Variable & Left state & Right state \\
\hline
  $\rho$ (g cm$^{-3}$)  & $1.7942$ &  $1.0$ \\
  $P$ (dyn cm$^{-3}$)   & $0.017942$ &  $0.01$ \\
  $\mathbf{u}$ (cm s$^{-1}$) & $-0.9759$ & $-1.751$ \\
                             & $-0.6561$ & $0.0$ \\
                             & $0.0$ & $0.0$ \\
  $\mathbf{B}$ (G)       & $1.0$ & $-1.0$ \\
                             & $1.74885$ & $0.6$ \\
                             & $0.0$ & $0.0$ \\
\hline
\end{tabular} 
\end{table}

Initially, the left state spans $x \in [0;0.3]$ while the region $x~\in~]0.3;1]$ is in the right state. We put the discontinuity at $x=0.3$ to let the fluid oscillate in the low density region. The boundaries are periodic in the $y$ and $z$-directions, and are fixed with the left and right states in the $x$-direction. The box has a $1$~cm size. We used the AMR in order to gain computation time while keeping a good precision in the sensitive zones. The minimum resolution is $1/64$ cm (level $\ell_\mathrm{min}=6$ of AMR), and we refine the grid to keep the magnetic field variation lower than $5\%$ in one cell, up to a resolution of $1/256$ cm (level eight of AMR).
We let the system evolve until the initial discontinuity converges toward a steady-state profile of the shock. The analytical calculation of the theoretical profile is detailed in Appendix \ref{shockanal}.

Figure \ref{shock} displays both the numerical steady-state profiles and the analytical solution. The agreement is very good with a difference lower than 1\% at shock interface. The larger errors in outer regions are oscillations due to the low resolution and the fixed boundary conditions, which happen even without the Hall effect.

\begin{figure*}
\begin{center}
\includegraphics[angle=0,width=0.9\textwidth]{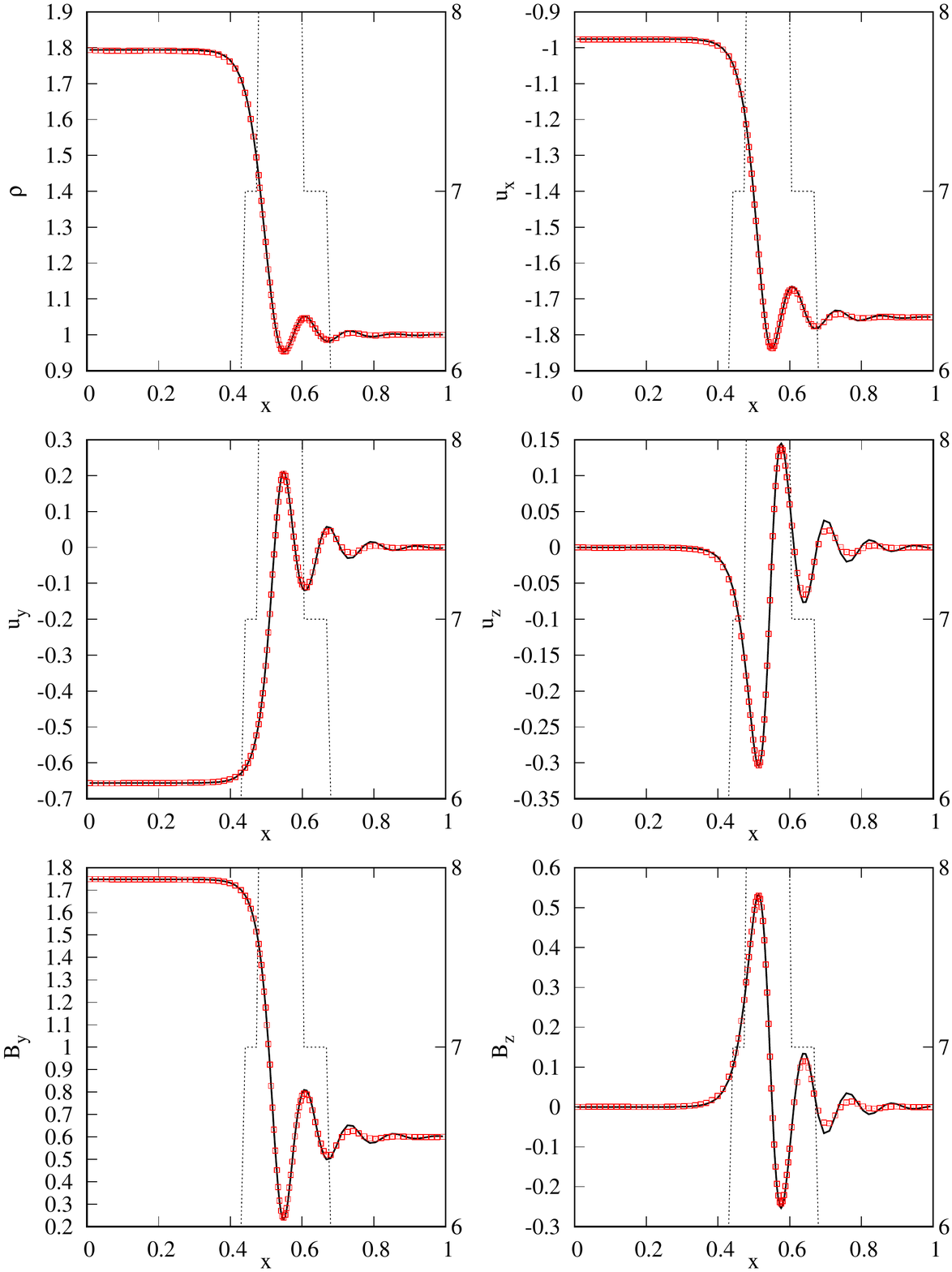}
\caption{Comparison between the analytical solution of the standing shock (solid black line) and the numerical result (red squares). The AMR level is represented as the dashed black line (right axis). From top to bottom: $\rho$ and $u_x$, $u_y$ and $u_z$, $B_y$ and $B_z$.}
\label{shock}
\end{center}
\end{figure*}

\section{Collapse of a dense core}\label{sect_collapse}

In this section, we describe our simulations of dense core collapse with the Hall effect as a practical test of our implementation. Simulations are presented in Sects \ref{sectSetup} and \ref{sectResults}, following elements of Hall effect theory in this context in \ref{Theory_hall_disk}.

\subsection{Hall effect in star formation: Theory} \label{Theory_hall_disk}

Assuming axi-symmetry and accounting only for the Hall effect, the azimuthal component of Equation \eqref{redind} reads

\begin{align}
  \frac{\partial B_\theta}{\partial t} =& - \left[\nabla \times \left(\frac{\eta_\mathrm{H}}{||\mathbf{B}||} \left(\nabla \times \mathbf{B}\right)\times \mathbf{B}\right)\right]_\theta \nonumber\\
                                       =& - \frac{\eta_\mathrm{H}}{B}\Bigg[\frac{\partial}{\partial z} \left[ B_z \left( \frac{\partial B_r}{\partial z} - \frac{\partial B_z}{\partial r} \right) - \frac{B_\theta}{r}\frac{\partial (rB_\theta)}{\partial r} \right] \nonumber \\
                                        &+ \frac{\partial}{\partial r} \left[ B_\theta \frac{\partial B_\theta}{\partial z} + B_r \frac{\partial B_r}{\partial z} + B_r \frac{\partial B_z}{\partial r} \Bigg] \right].
\end{align}
This expression shows that Hall effect generates a toroidal magnetic field $B_\theta$, provided that its radial and vertical components $B_r$ and $B_z$ are not uniform. 

During the collapse of a magnetized dense core, the pinching of the field lines creates an accumulation of matter known as the "pseudo-disk" \citep{GalliShu93b}, with a density typically higher than $10^{-15}$ g cm$^{-3}$. It is perpendicular to the largescale magnetic fields and is located where the pinching is the strongest, at the neck of the "hourglass" formed by the field lines. The large spatial variations of the magnetic field at this location generates a strong toroidal electric current in the mid-plane of the pseudo-disk. $\mathbf{u}_\mathrm{H}$ being directly proportional to $\mathbf{J}$, the field lines are then twisted in the plane of the pseudo-disk. If the core is initially rotating, the fluid also twists the lines. The Hall effect therefore strengthens or weakens the toroidal magnetic field generated by the rotation, depending on the sign of $-\eta_\mathrm{H} J_\theta$. The consequence is a stronger or weaker magnetic braking, resulting in a slower or faster rotation speed of the fluid \citep[see e.g.,][]{Tsukamoto2015}. If the core is not rotating initially, the Hall effect still generates a toroidal magnetic field, which induces the rotation of the fluid, in the same way that the magnetic braking slows the rotation via the magnetic tension. In either case, the fluid is accelerated in the direction of $- \mathbf{u}_\mathrm{H}$.

\subsection{Setup}\label{sectSetup}

The objective is to test our implementation with a simple core-collapse simulation, and study the role of the Hall effect.
We used the standard \citet{1979ApJ...234..289B} initial conditions with a non-rotating uniform sphere of gas. The sphere radius is $3712$ au and its density $\rho_0=4.15 \times 10^{-18}$ g cm$^{-3}$ for a total mass of $1.5$ M$_\odot$. The thermal over gravitational energy ratio is $\alpha = 0.25$ with a uniform temperature $T=10$ K. The external medium is $100$ times less dense. The magnetic field is uniform at $B=90.3~\mu G$ in the whole box along the $z$-direction, which is equivalent to a mass-to-flux ratio of $\mu=7$. 
We solved the complete set of Equations \eqref{testmass}-\eqref{testenergy} without considering the ambipolar and the Ohmic diffusions. The Hall resistivity is set at a fixed value of $\eta_\mathrm{H} = 10^{20}~$cm$^2~$s$^{-1}$. The length of the simulation box is four times the sphere radius and we use the generalized moncen slope limiter with a $1.5$ coefficient.
This setup is now refereed as our "reference case".
To mimic the evolution of temperature during the collapse up to the first Larson core, we used the following barotropic equation of state

\begin{equation}
  T=10 \left[ 1+\left(\frac{\rho}{10^{-13}~\mathrm{g}~\mathrm{cm}^{-3}} \right)^{\frac{5}{3}} \right] .
\end{equation}

The initial grid is refined at AMR level $5$ ($32^3$) and we imposed at least eight points per Jeans length to satisfy the \citet{1997ApJ...489L.179T} condition of at least four points per Jeans length. The maximum resolution is $\Delta x=2$ au at the end of the simulation.
The "formation of the first core" is the time $t_\mathrm{C}$ at which the maximum density reaches $10^{-13}$ g cm$^{-3}$.
Finally, we refer to a structure rotationally supported against gravity as "disk". We used the criteria of \citet{joos}
\begin{itemize}
  \item[-] density criterion: $\rho > 3.8 \times 10^{-15}$ g cm$^{-3}$,
  \item[-] rotationally supported: $u_\theta > 2 u_r$, with $u_\theta$ and $u_r$ the azimuthal and radial velocities,
  \item[-] hydrostatic equilibrium within the disk: $u_\theta > 2 u_z$,
  \item[-] not within the first core: $\rho u^2/2 > 2P_\mathrm{therm}$, with $P_\mathrm{therm}$ the thermal pressure.
\end{itemize}

\subsection{Results}\label{sectResults}

\begin{figure}
\begin{center}
\includegraphics[trim= 5cm 3cm 5cm 3.2cm, width=0.49\textwidth]{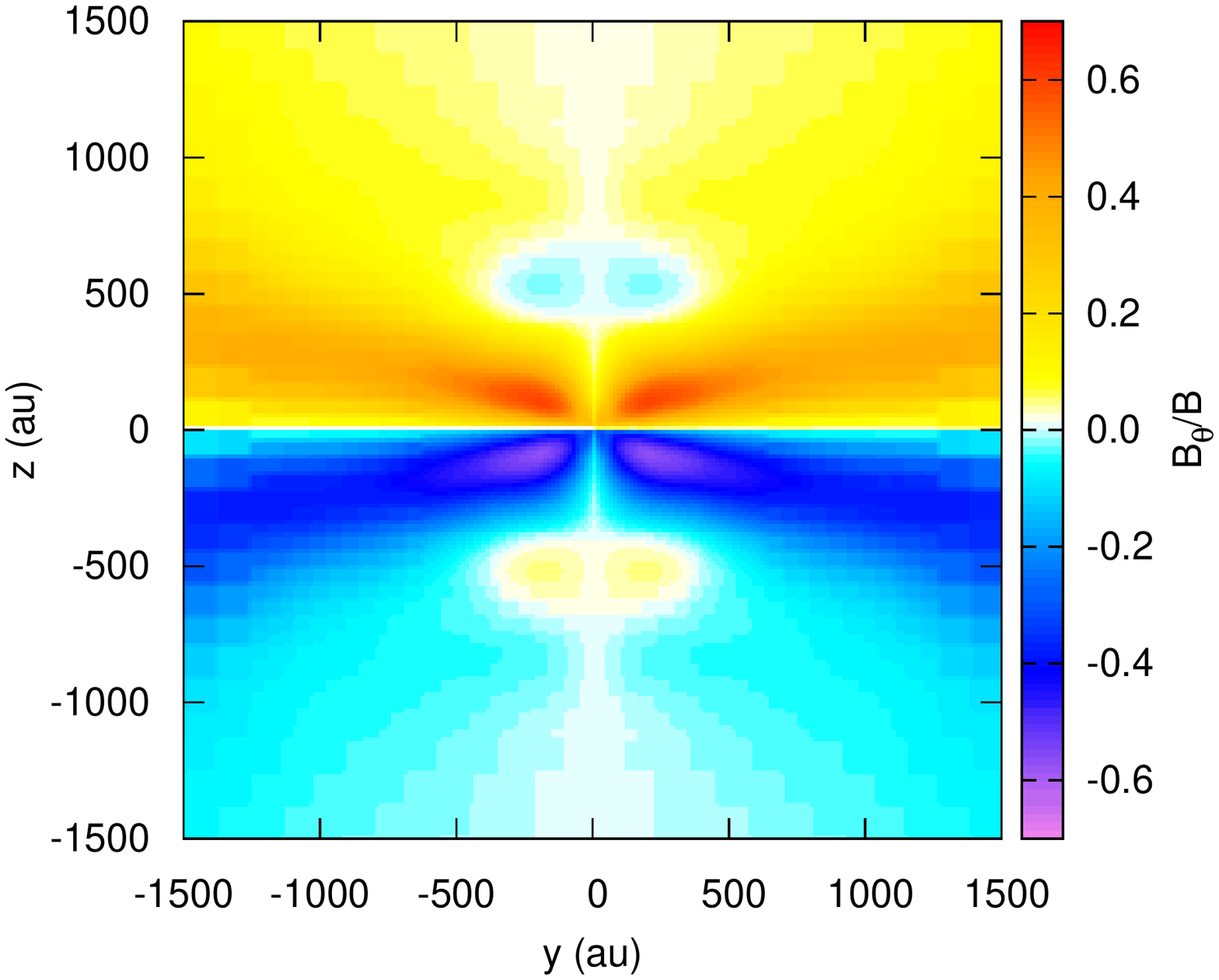}
\includegraphics[trim= 5cm 2cm 5cm 3cm, width=0.49\textwidth]{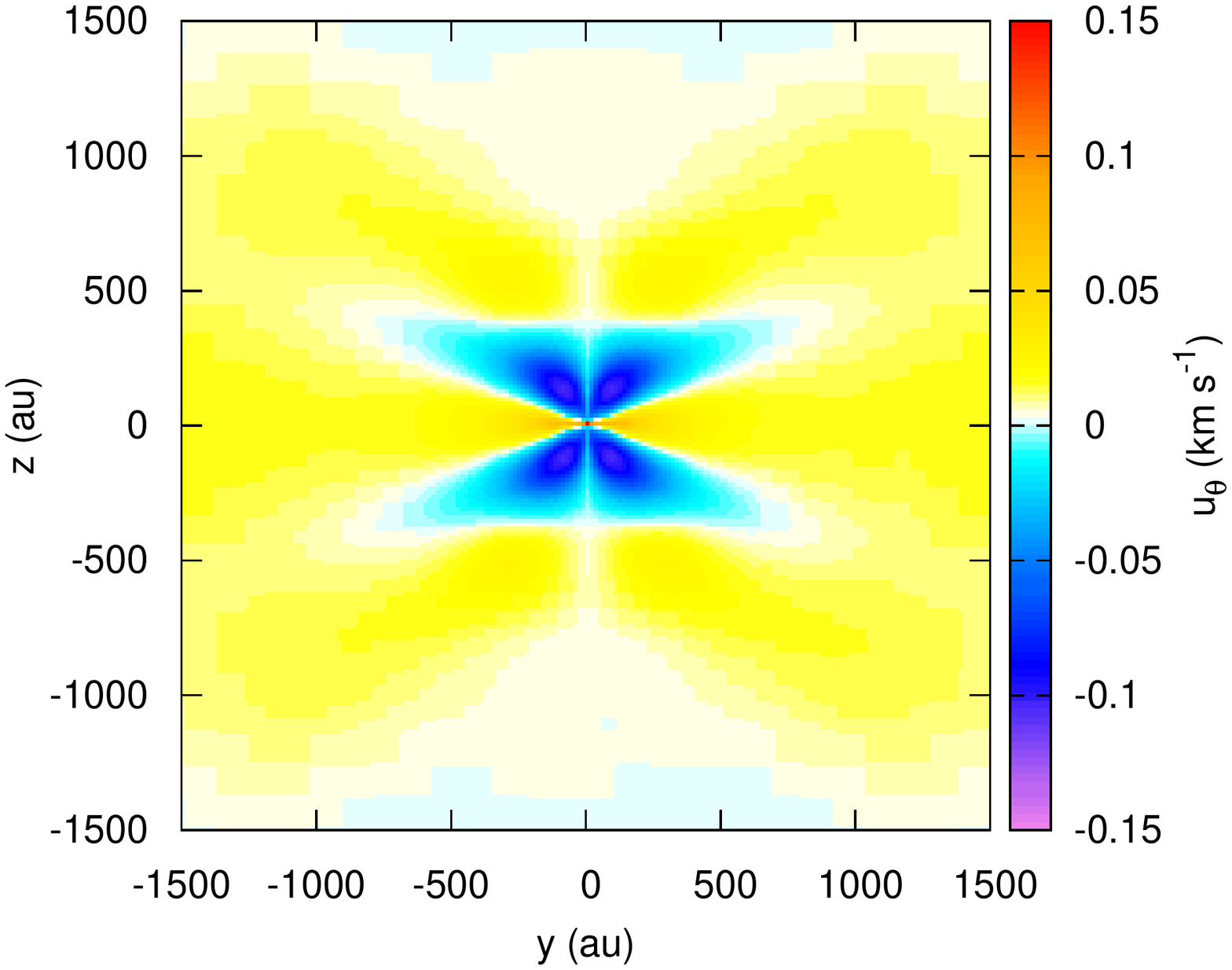}
\caption{Toroidal magnetic field $B_\theta$ (top) and azimuthal velocity $u_\theta$ (bottom) on the plane $x=0$, $0.230$ kyr after the formation of the first core.}
\label{bphi}
\end{center}
\end{figure}

A differential rotation of the cloud starts about an axis in the same direction as the initial magnetic field, with an increased amplitude over time. Figure \ref{bphi} shows maps on the plane $x=0$ of the ratio of the toroidal magnetic field over the total magnetic field amplitude $B_\theta/B$ and of the azimutal velocity $u_\theta$, $0.230$ kyr after the formation of the first core.
The twisting of the field lines is the strongest 200 au above and below the mid-plane. A large quantity of angular momentum is created in the disk while the whistler waves transport its negative counter-part along the field lines. This negative angular momentum is transferred to the regions above and below the disk, thus setting a backward rotation that is visible in blue colors in the bottom panel. The momentum of the core and disk are therefore compensated by the counter-rotation of an hourglass-shaped envelope of $1000$~au in height and diameter. With a similar setup, \citet{2011ApJ...733...54K} and \citet{LiKrasnopolskyShang} obtain the same qualitative results, with counter-rotating regions in the envelope. 

We then compared the azimuthal velocity $u_\theta$ of the fluid to the azimuthal Hall speed and to the Keplerian velocity. The Hall speed is defined accordingly to Section \ref{principle}
\begin{equation}
  \mathbf{u}_\mathrm{H} = \eta_\mathrm{H} \frac{\mathbf{J}}{\|\mathbf{B}\|}.
\end{equation}
Using the second Newton law, the Keplerian velocity was calculated as
\begin{equation}
  u_\mathrm{Kepler}(\mathbf{r}) = \left[r \left(\sum_{\substack{i=1,\mathrm{ncells} \\ \mathbf{r}_i \neq \mathbf{r}}}\frac{Gm_i}{\|\mathbf{r}_i-\mathbf{r}\|^3}(\mathbf{r}_i-\mathbf{r})\right)_r \right]^\frac{1}{2}.
\end{equation}
for each cell in our discretized system (assuming the center of mass is at the origin).

\begin{figure}
\begin{center}
\includegraphics[trim= 2cm 1cm 2cm 1cm, width=0.49\textwidth]{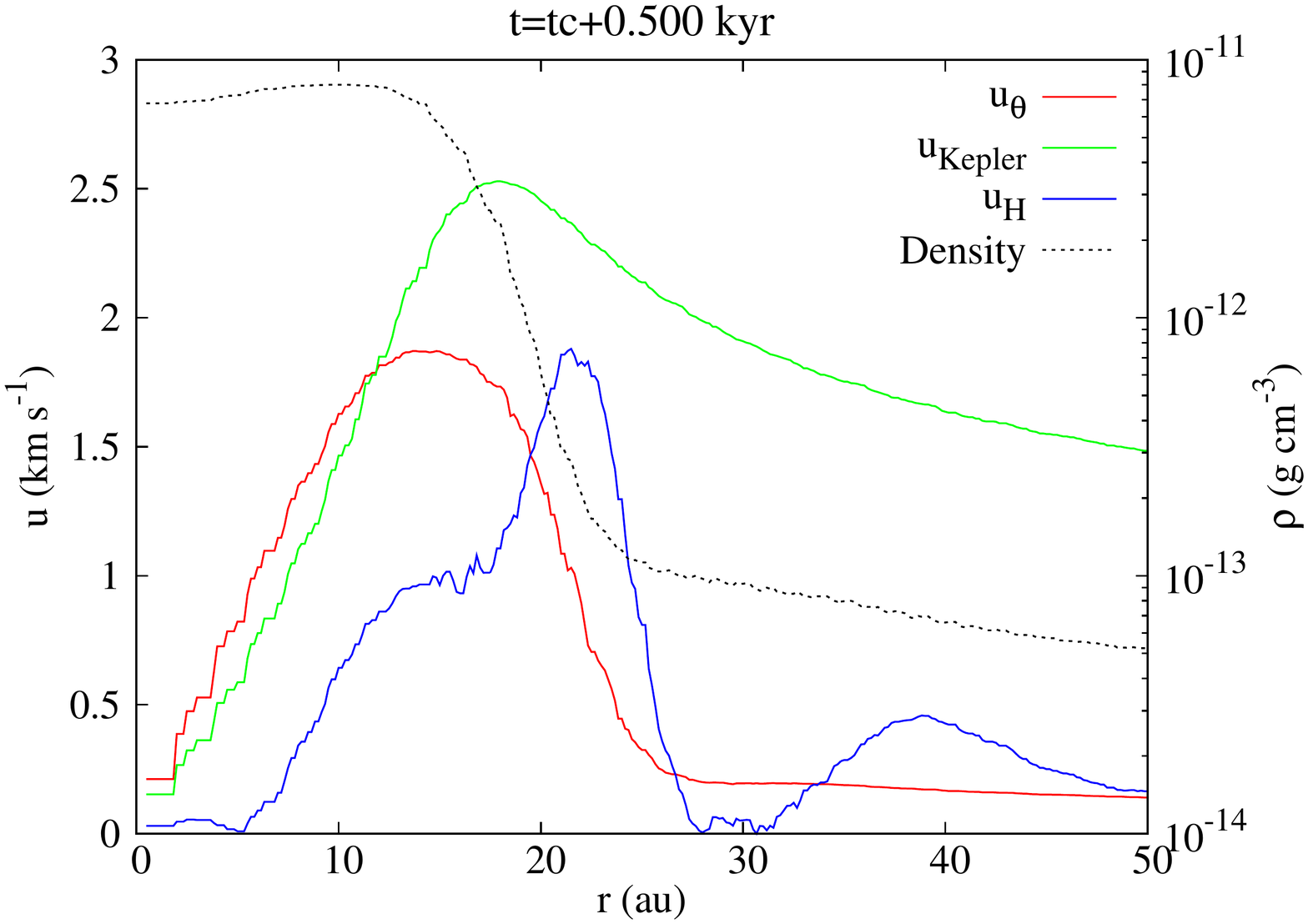}
\includegraphics[trim= 2cm 1cm 2cm 1cm, width=0.49\textwidth]{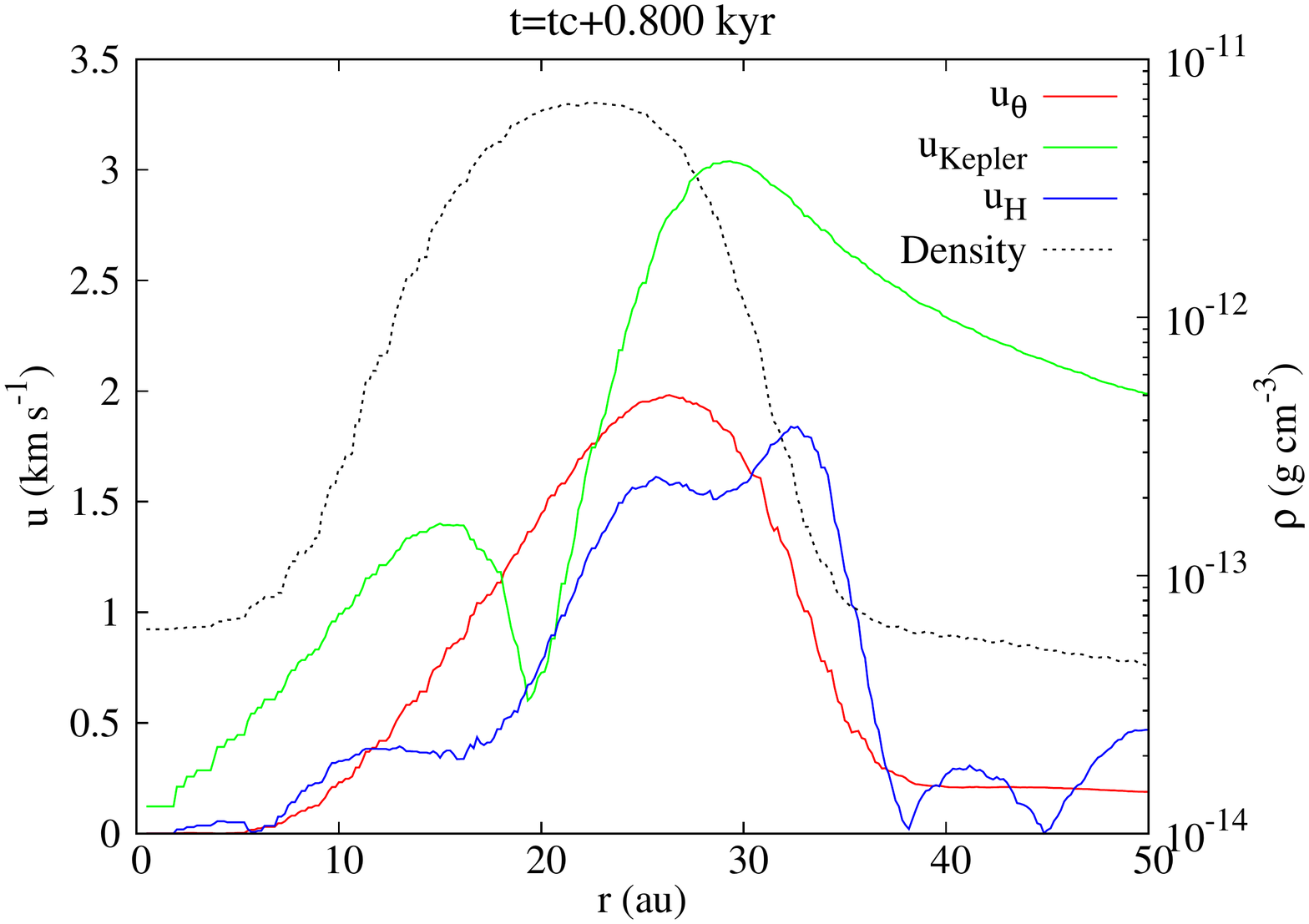}
  \caption{Radial profiles ($z=0$, azimuthal averaging) of the azimuthal velocity (red curve), the azimuthal Hall speed (blue curve) and the Keplerian velocity (green curve) at $t=t_\mathrm{c}+0.500$ and $t=t_\mathrm{c}+0.800$ kyr (top and bottom panel). The black dotted line represents the density profile (right axis).}
\label{vhallvkepvtheta}
\end{center}
\end{figure}

\begin{figure}
\begin{center}
\includegraphics[trim= 2cm 1cm 2cm 1cm, width=0.49\textwidth]{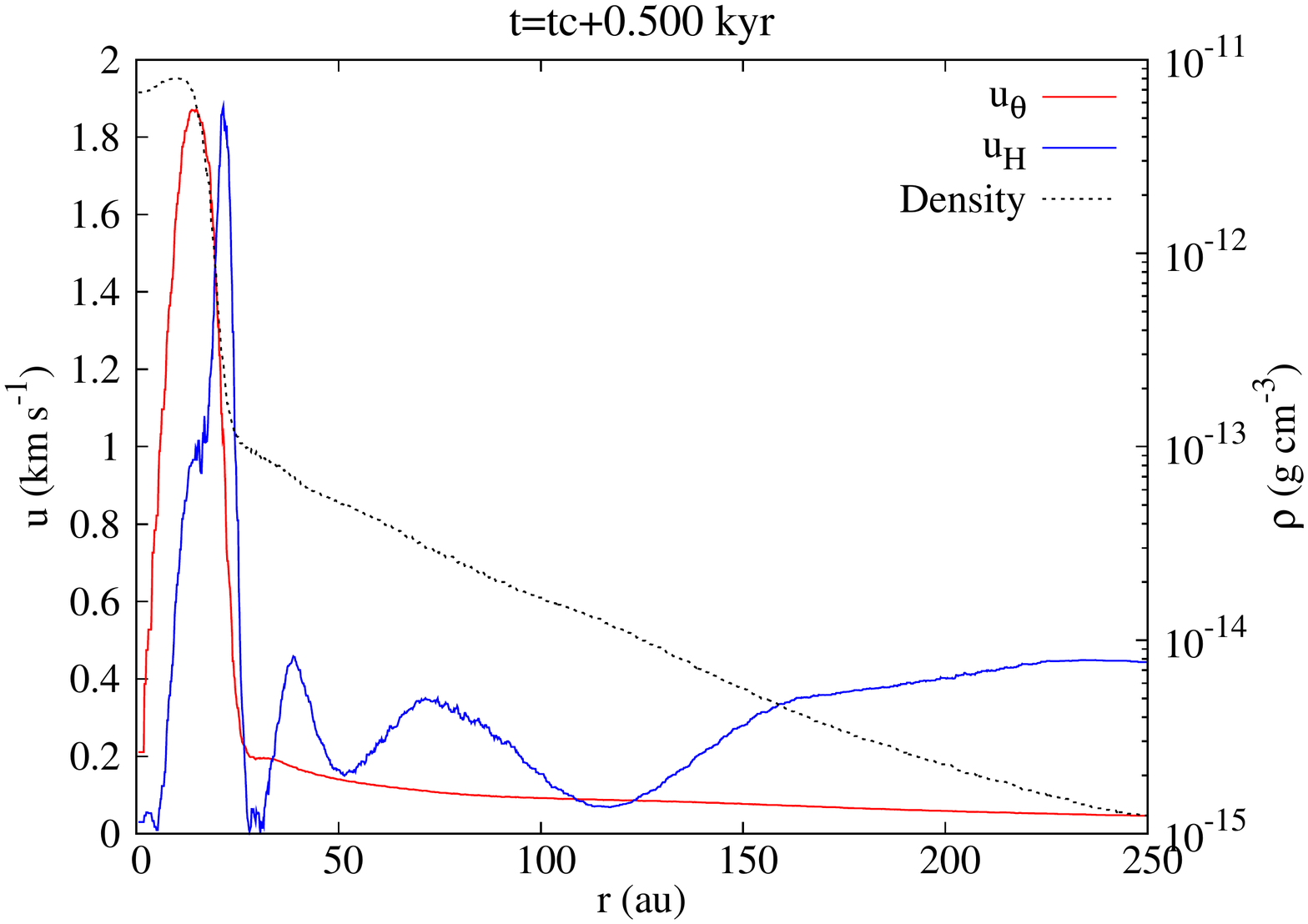}
\includegraphics[trim= 2cm 1cm 2cm 1cm, width=0.49\textwidth]{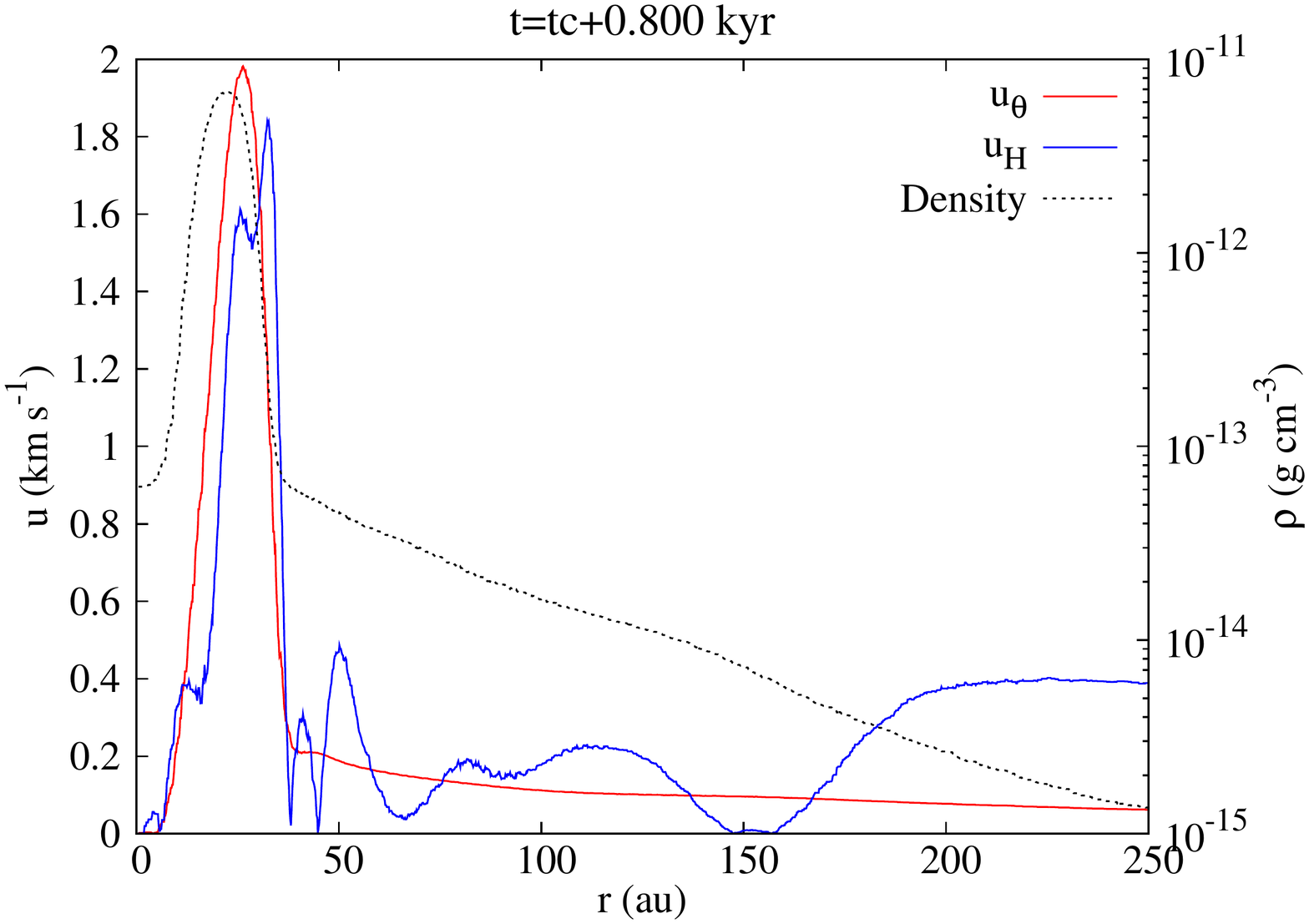}
\caption{As Figure \ref{vhallvkepvtheta}, at larger radius and without the Keplerian velocity.}
\label{vhallplusloin}
\end{center}
\end{figure}

\begin{figure}
\begin{center}
\includegraphics[trim= 6cm 2cm 5cm 3cm, width=0.49\textwidth]{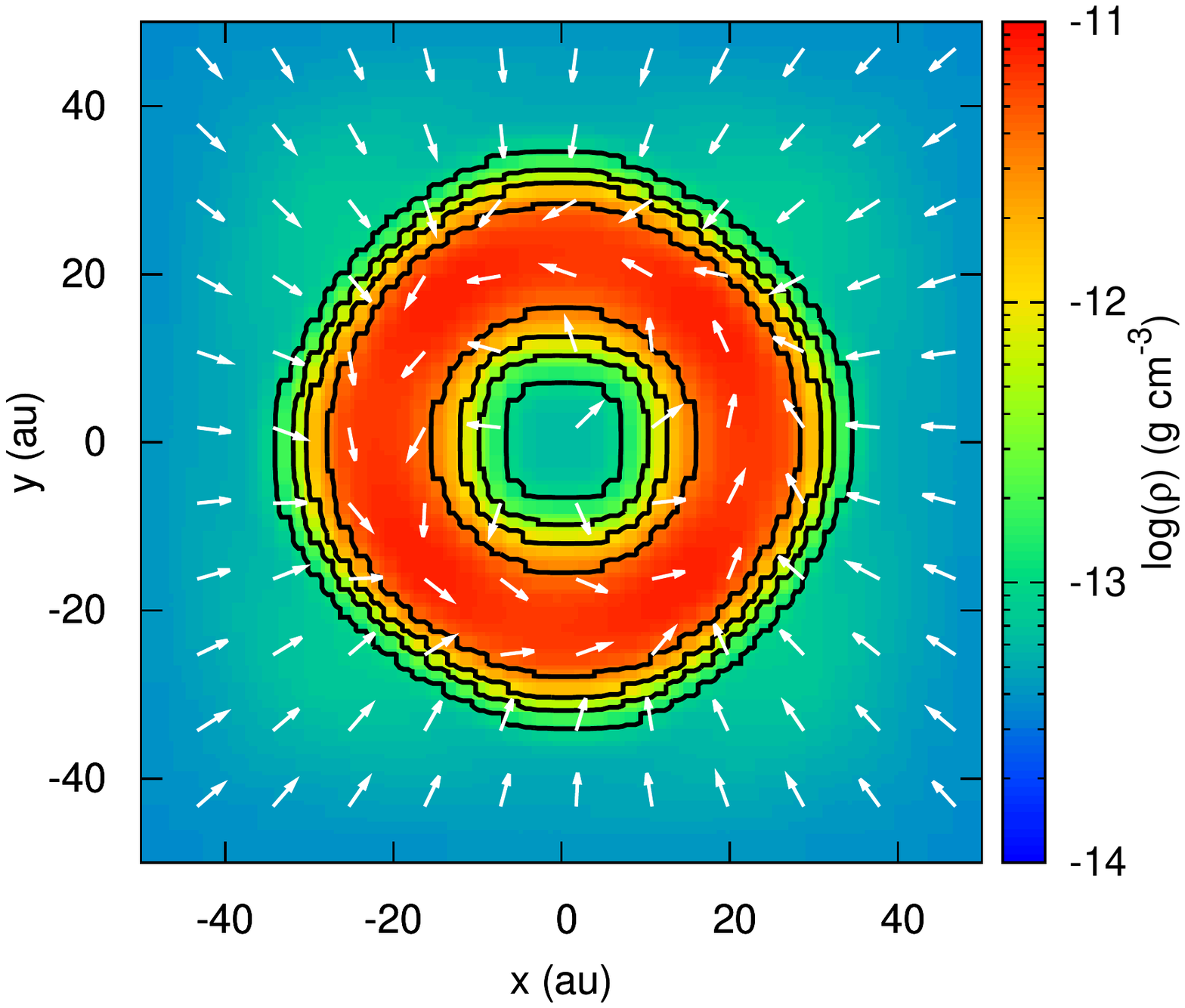}
  \caption{Density and velocity map of the reference case at $t=t_\mathrm{C}+0.800$ kyr.}
\label{rhov}
\end{center}
\end{figure}

Figure \ref{vhallvkepvtheta} shows the radial profile of the three velocities $0.500$ and $0.800$ kyr after the first core formation. Figure \ref{vhallplusloin} is a zoom-out of Figure \ref{vhallvkepvtheta} without the Keplerian velocity. All radial profiles are azimuthally averaged.
In the inner regions, the fluid rotation speed is equal to the Keplerian velocity and larger than the Hall speed. Eventually, the centrifugal force overcomes gravity, which results in a positive radial velocity of the fluid and the formation of a ring-shaped core, as shown in Figure \ref{rhov}. All the velocities drop outside the Larson core, beyond $30$ au, and the fluid is roughly rotating at the Hall velocity, as seen in Figure \ref{vhallplusloin}. The two values do not perfectly match but differ by less than $50$\%. In the absence of every other forces but the Lorentz force, the two velocities should correspond, but in our case, the presence of gravity and thermal pressure prevent a perfect match. The coupling is effective up to $r\approx 160$~au, down to a density of $\rho\approx 10^{-14}$ g cm$^{-3}$. As explained in Section \ref{Theory_hall_disk}, the region we could define as the pseudo-disk is rotating at the Hall velocity. Around $30$ au at $t=t_\mathrm{c}+0.500$ kyr and $40$ au at $t=t_\mathrm{c}+0.800$ kyr, the azimuthal Hall velocity changes sign in a small interval. This component of $u_\mathrm{H}$ is proportional to $\mathbf{J}_\theta = \frac{\partial B_r}{\partial z} - \frac{\partial B_z}{\partial r}$. While $B_z$ decreases with r, there is a small interval at the outer edge of the core, before the accretion shock, in which its decrease is much lower. As a result, $\frac{\partial B_z}{\partial r}$ is lower (in negative value) while $\frac{\partial B_r}{\partial z}$ keeps an approximately constant value, causing a temporary change of sign.

\begin{figure}
\begin{center}
\includegraphics[trim= 1cm 1cm 1cm 1cm, angle=0, width=0.49\textwidth]{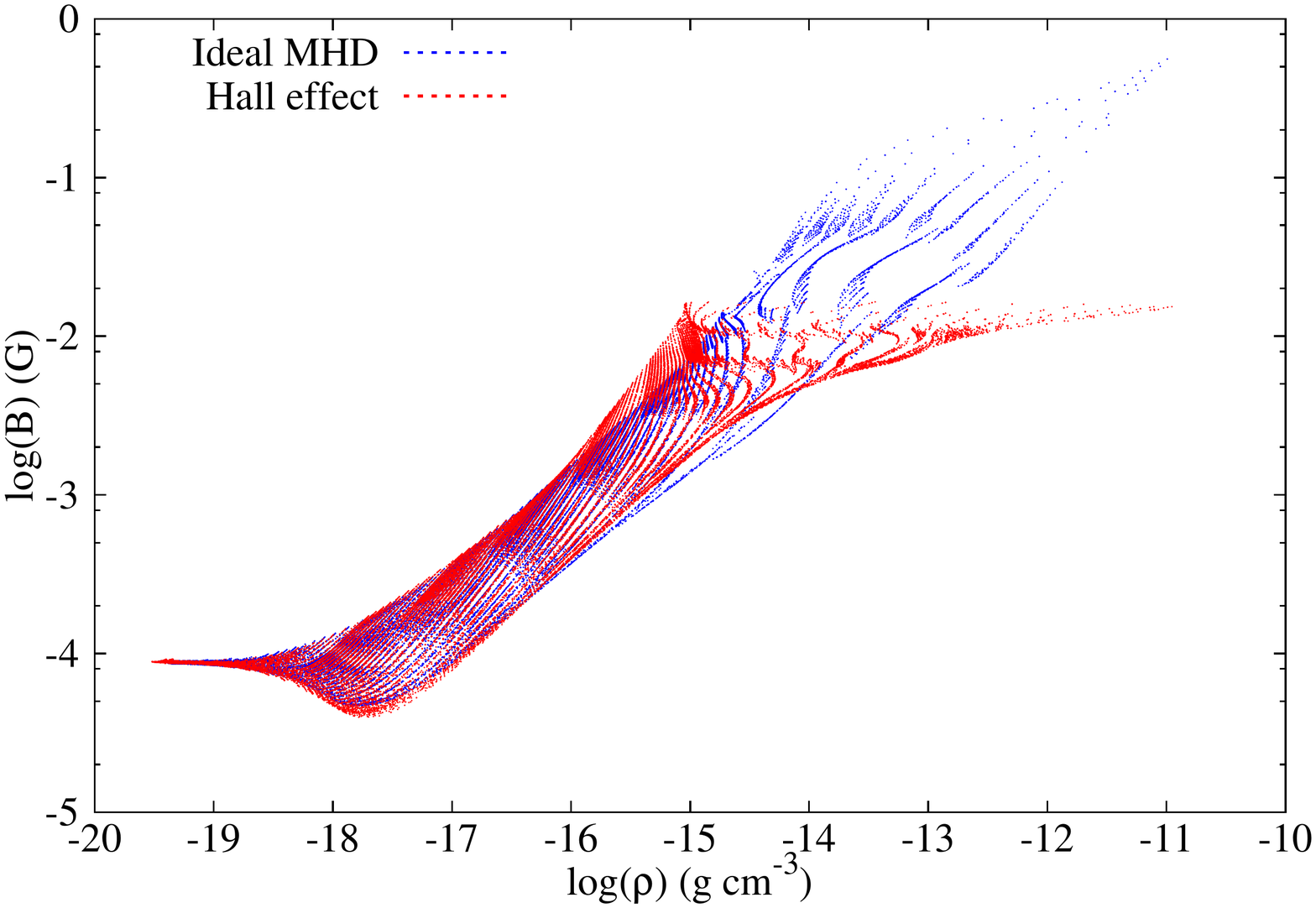}
  \caption{Magnetic field amplitude as a function of the density for two simulations with and without Hall (red and blue domains respectively). The snapshot is taken when the maximum density reaches $5 \times 10^{-12}$~g~cm$^{-3}$ (at $t=t_\mathrm{C}+0.220$ kyr for the Hall simulation).}
\label{brho}
\end{center}
\end{figure}

The Hall effect redistributes the angular momentum within the envelope, and we know adress the case of magnetic flux. Figure \ref{brho} shows the magnetic field amplitude as a function of the density for each cell in the box for this simulation and its counter-part without the Hall effect (i.e., ideal MHD), when simulations both reach a maximum density of $\rho = 5\times 10^{-12}$~g~cm$^{-3}$. The magnetic field in ideal MHD is roughly $0.5$~G at the center, while it is bounded to $3 \times 10^{-2}$~G with the Hall effect. On the contrary, the magnetic field amplitudes are very similar at densities below $10^{-15}$-$10^{-14}$ g cm$^{-3}$, confirming that the Hall effect becomes relevant above these densities. \citet{DA1} found a comparable "plateau" using the ambipolar diffusion, which also redistributes the magnetic flux and yields a similar profile for the magnetic field. Both non-ideal MHD effects then act to regulate the magnetic flux.

\subsection{The issue of angular momentum conservation}

\subsubsection{Presentation of the problem}

Here we introduce some notations that are useful to the remainder of the discussion. We note $\mathbf{L}$ the angular momentum vector
\begin{equation}
  \mathbf{L}= \sum_{i=1,\mathrm{ncell}} m_i\mathbf{r}_i \times \mathbf{u}_i,
\end{equation}
where $m_i$, $\mathbf{r}_i$ and $\mathbf{u}_i$ represent the mass, the position vector and the velocity vector of cell $i$.
The system being nearly axi-symmetric, the main component of this vector is $L_z$ (i.e., the initial magnetic field direction), that can also be calculated as
\begin{equation}
  L_z = \sum_{i=1,\mathrm{ncell}} m_i r_{\mathrm{c},i} u_{\theta,i},
\end{equation}
with $r_\mathrm{c}$ the cylindrical radius and $u_\theta$ the azimuthal velocity. We define the "positive" and "negative" momenta $L_z^+$ and $L_z^-$ as follows

\begin{align}
  L_z^+ &= \sum_{i, v_{\theta,i} > 0} m_i r_{\mathrm{c},i} u_{\theta,i}, \\
  L_z^- &= \sum_{i, v_{\theta,i} < 0} m_i r_{\mathrm{c},i} u_{\theta,i}. \\
\end{align}
We therefore have $L_z^+ +L_z^- = L_z \cong \|\mathbf{L}\|$.

In our setup, there is initially no angular momentum in the system. $L_z^+$ thus represents the "positive" rotation created by the Hall effect, balanced by the "negative" rotation $L_z^-$, and the numerical solution should verify $\|L_z^+\| \approx \|L_z^-\|$. Usually, $L_z$ does not exactly equal zero because of truncation errors, but this is not an issue if $L_z$ remains small (typically, compared with $L_z^+$ and $L_z^-$).

\begin{figure}
\begin{center}
\includegraphics[trim= 2cm 1cm 2cm 1cm, width=0.49\textwidth]{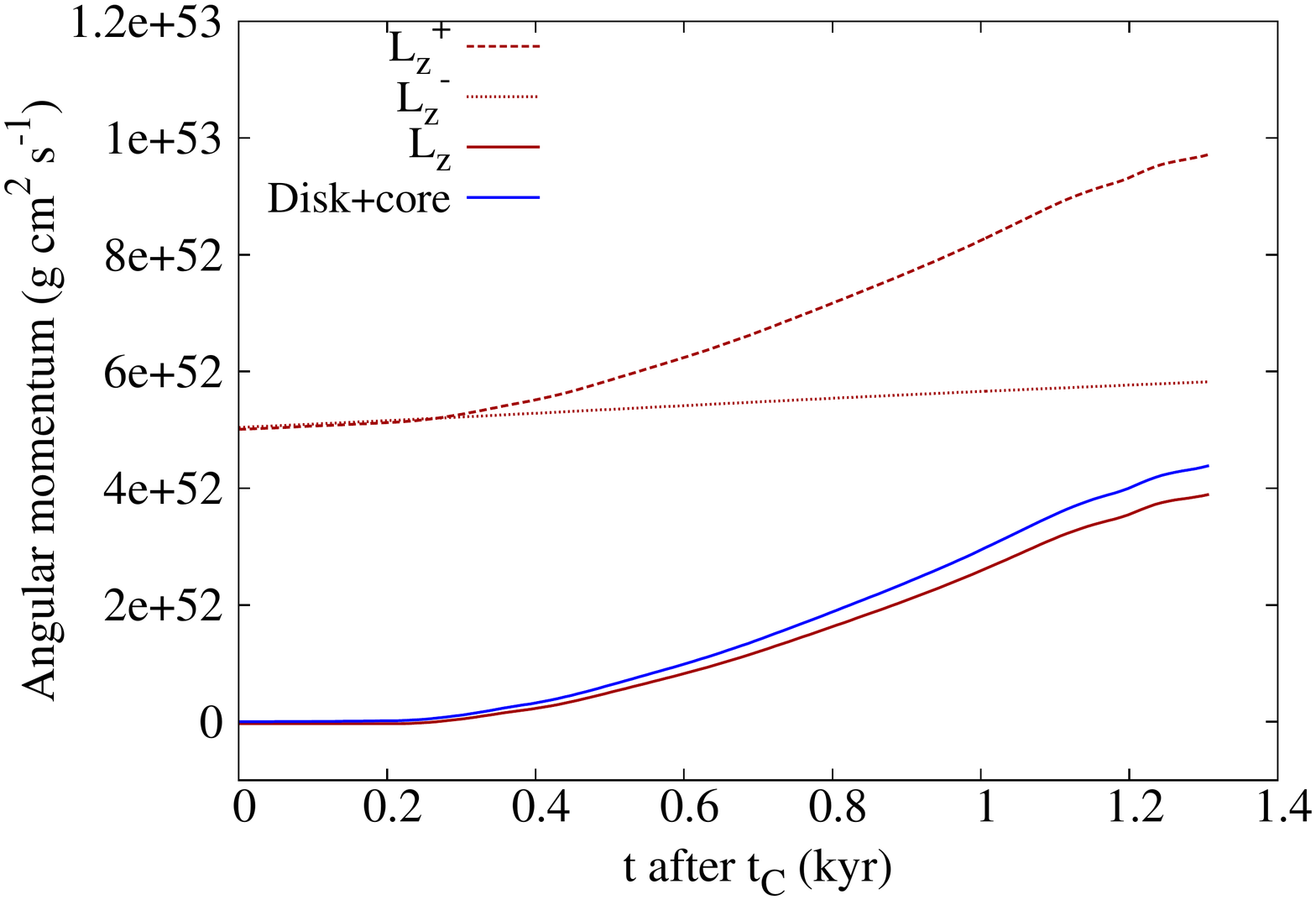}
  \caption{Positive $L_z^+$ and negative $L_z^-$ angular momentum evolution after the formation of the first Larson core, in red dashed and dotted line respectively. The red solid line shows the total angular momentum $L_z$, and the blue line the angular momentum of the core+disk region.}
\label{momhall}
\end{center}
\end{figure}

In Figure \ref{momhall}, we have plotted the evolution of $L_z^+$, $L_z^-$ and $L_z$ during the simulation after the formation of the first Larson core. In the remainder of the manuscript, this reference case is plotted in red.
From $t=t_\mathrm{c}+0.230$ kyr, the positive momentum skyrockets and almost doubles its value by $t=t_\mathrm{c}+1.300$ kyr, while the negative momentum increases by less than 10\%. The total angular momentum $L_z$ then increases as well and reaches $2/3$ of the value of $L_z^-$. The angular momentum is therefore not conserved in the simulation box. 

\citet{2011ApJ...733...54K} also encounter a non-conservation of angular momentum in their simulation with a similar setup. They attribute this phenomenon to torsional Alfv\'en waves leaving the simulation box. It is however not the case here since the boundary conditions are periodic. The blue curve of Figure \ref{momhall} represents the angular momentum of the disk and the core, which we define as the gas with density $\rho > 10^{-13}$ g cm$^{-3}$. The disk+core angular momentum shows the same {\bf time} evolution as the excess of angular momentum created in the simulation box, with $L_z \approx 0.85 L_\mathrm{core+disk}$. This strongly suggests that the problem takes its origin in the central region, and we have calculated that whistler waves do not have time to travel from the center to the outskirt of the box. The same issue arises for different boundary conditions, box sizes, and slope limiters (see Appendix \ref{app_slopebound}).

\subsubsection{Whistler waves on discretized space}

\begin{figure}
\begin{center}
\includegraphics[trim= 2cm 1cm 1cm 1cm, width=0.49\textwidth]{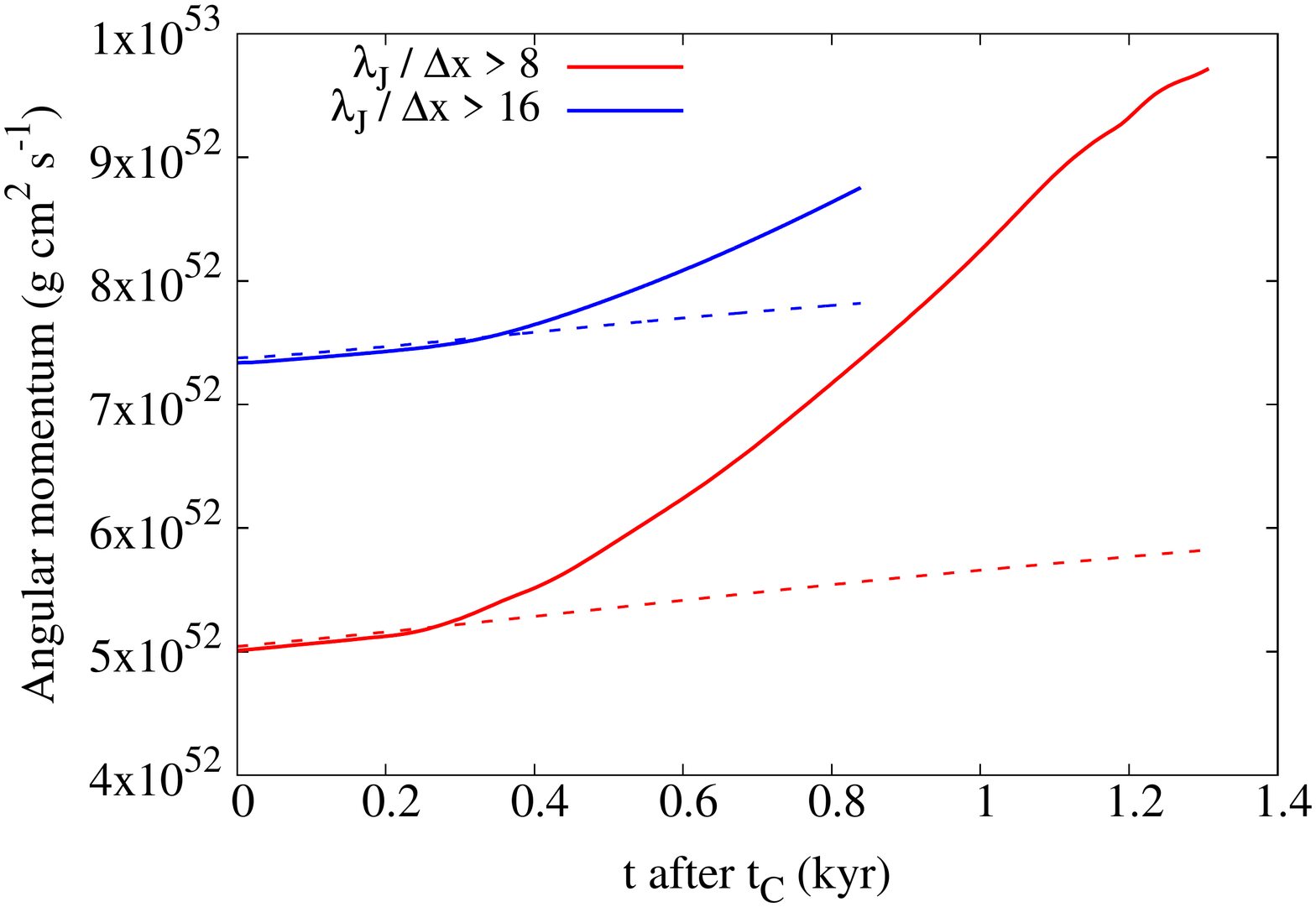}
\includegraphics[trim= 2cm 1cm 1cm 1cm, width=0.49\textwidth]{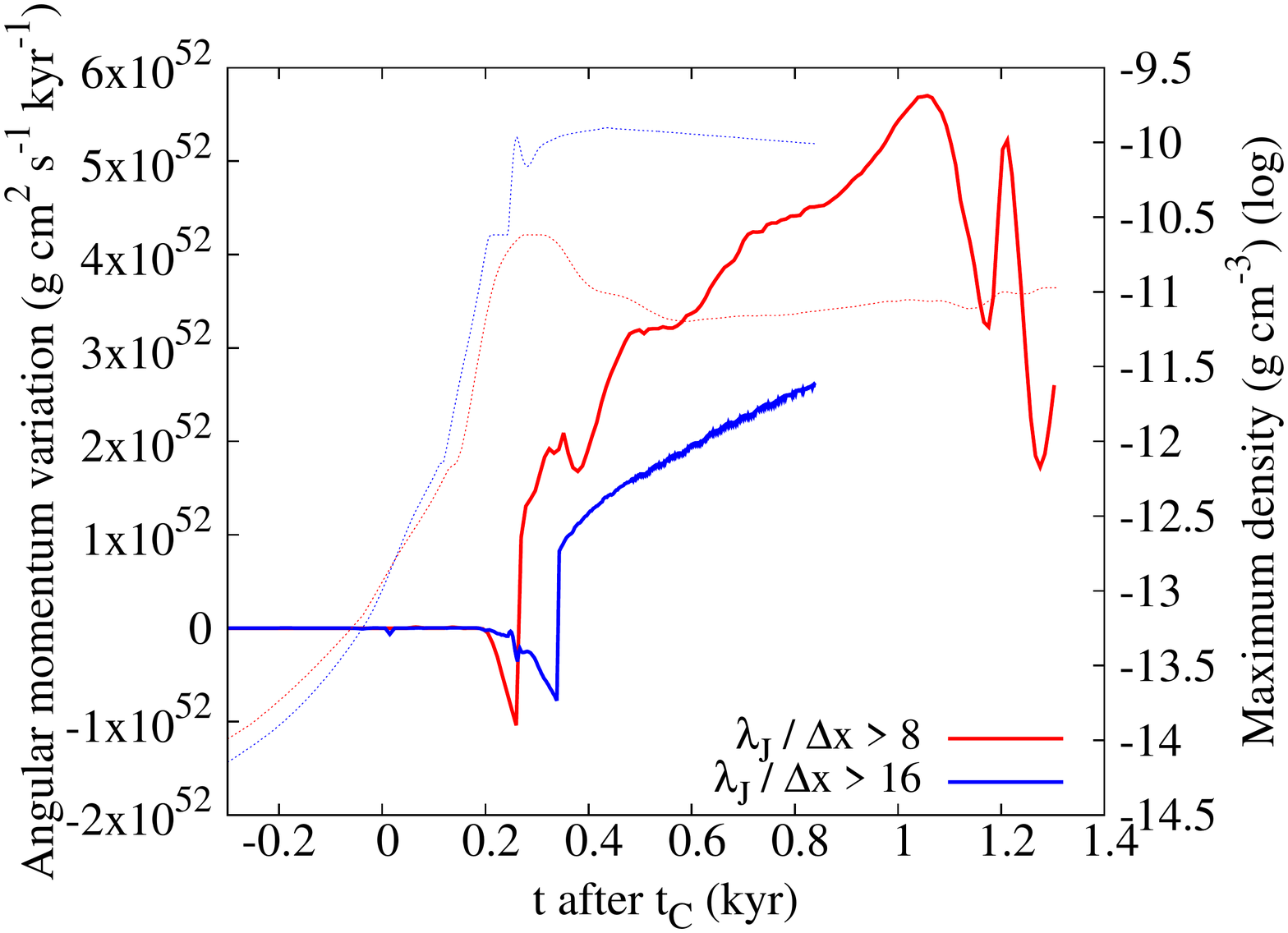}
  \caption{Top panel: Angular momentum evolution of the two simulations with different refinement criteria (Red: $\lambda_\mathrm{J}/\Delta x > 8$, Blue: $\lambda_\mathrm{J}/\Delta x > 16$). The solid curves represent $L_z^+$ and dashed curves represent $L_z^-$. Bottom panel: Temporal variation of the positive angular momentum in solid lines, and evolution of the maximum density in dotted lines.}
\label{angmom_jeans}
\end{center}
\end{figure}

We investigated the role of the spatial resolution in this problem. In the previous run, the refinement criterion ensures at least 8 cells per local Jeans length ($\lambda_\mathrm{J}/\Delta x > 8$). We ran another simulation with a doubled Jeans length resolution criterion ($\lambda_\mathrm{J}/\Delta x > 16$). Figure \ref{angmom_jeans} shows the comparison between the angular momenta in absolute value for the two runs (top panel) as well as their temporal variation (bottom panel). While the most resolved simulation shows momenta larger by 50\% (probably because the Hall effect can act at smaller scales during the isothermal stage of the collapse), the increase of angular momentum is lower by roughly $50$\% in the most resolved case. The local maximum of the angular momentum variation, for the lower resolution case at $t=t_\mathrm{C}+0.400$ kyr, corresponds to the time at which the core expands due to the centrifugal force, as seen in the evolution of maximum density in dotted line. This "bouncing" also happens at a smaller scale for the $\lambda_\mathrm{J}/\Delta x > 16$ simulation at $t=t_\mathrm{c}+0.300$ kyr, but is tempered by the lower excess of angular momentum and the additional gravitational force (due to the higher density).

\begin{figure}
\begin{center}
\includegraphics[trim= 2cm 1cm 1cm 1cm, width=0.49\textwidth]{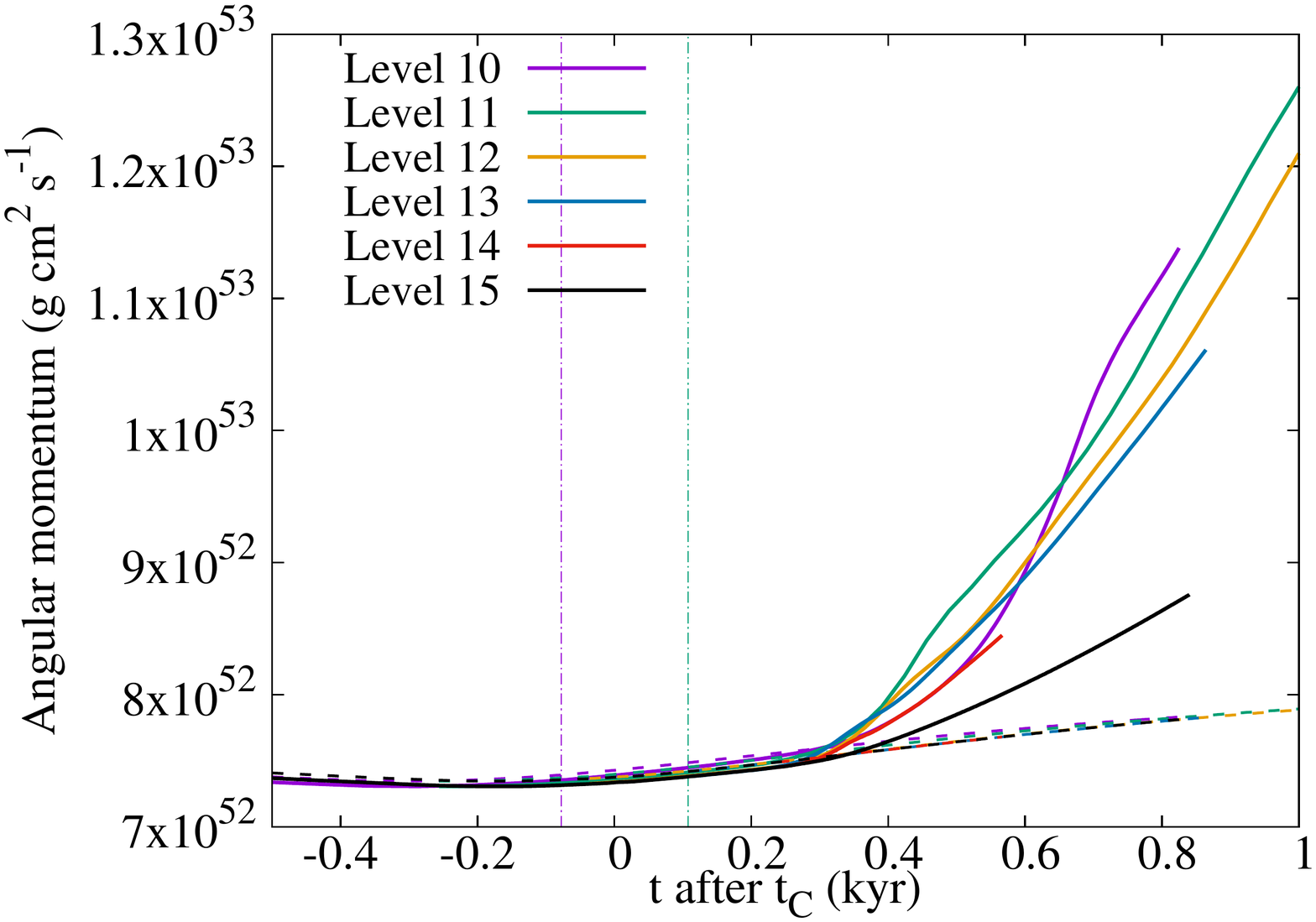}
  \caption{As top panel of Figure \ref{angmom_jeans}, but comparing different maximum resolutions. The vertical bars show when the \citet{1997ApJ...489L.179T} condition is no longer verified for maximum resolution levels of 10 and 11. Please note that the level 15 line in black is the same as the blue curve in Figure \ref{angmom_jeans}.}
\label{angmom_resol}
\end{center}
\end{figure}

We observed the same behavior when we limit the maximum resolution. In the following set of simulations, we use the 16 points per Jeans length criterion, but we did not allow the code to refine above a given level. Figure \ref{angmom_resol} represents the angular momentum evolution for maximum levels from $10$ ($\Delta x=16$ au) to $15$ ($\Delta x=0.5$ au). The increase of the angular momentum starts at the same time after the formation of the first Larson core for every simulation. The local maximum is lower and happens sooner as the resolution increases. This trend might be explained by the increase of the angular momentum starting in the most refined cells. With a lower resolution, cells are larger, and therefore hold more mass for the same density. Thus, the gas they harbor takes more time to be set in rotation due to inertia, and once it reaches the critical rotation speed at which the centrifugal force dominates gravity, there is more mass in rotation at a larger effective distance from the center of the simulation (because flow variables are computed at the cell centers), meaning a larger angular momentum. Nonetheless, a higher resolution seems to temper the increase.

\begin{figure}
\begin{center}
\includegraphics[trim= 2cm 1cm 1cm 1cm, width=0.49\textwidth]{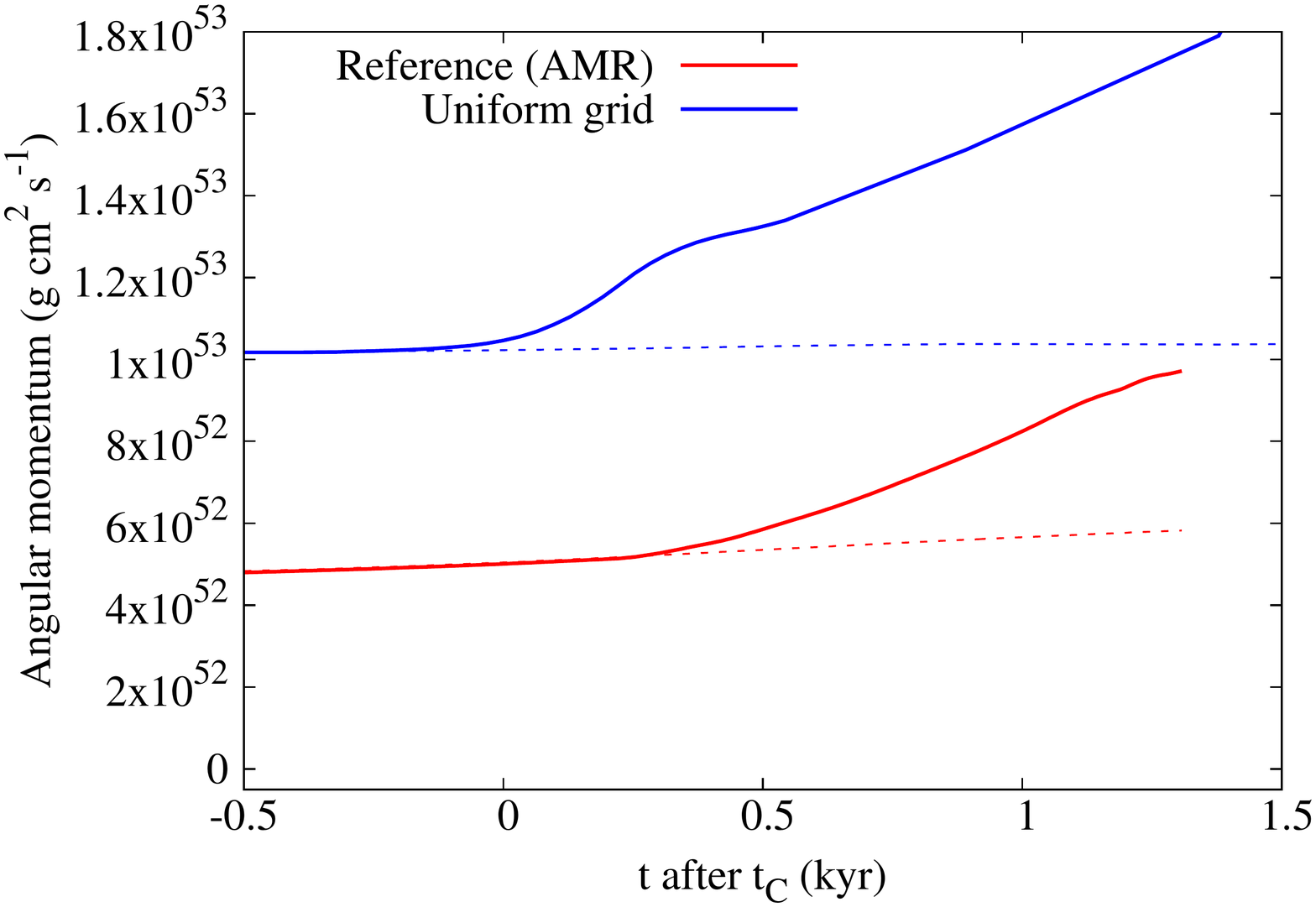}
  \caption{As top panel of Figure \ref{angmom_jeans} but comparing the reference AMR case (red) and the uniform 256$^3$ grid (blue).}
\label{angmom_lvl8}
\end{center}
\end{figure}

The dissipation of whistler waves on a discretized space could be at stake. As consequence of the AMR method, short and fast whistler waves generated in high resolution regions could simply be dissipated at refinement level interfaces due to a lack of resolution (the Shannon theorem). To test this hypothesis, we ran the same simulation with a uniform grid of 256$^3$ (level eight of refinement, or a resoluton of $64$ au). The aim is purely the numerical study since such a low resolution can not provide meaningful physical results, and a better uniform resolution would be prohibitively expensive. 
Figure \ref{angmom_lvl8} displays the angular momentum of this simulation compared to the reference case. The increase of angular momentum occurs at a greater intensity, and at an earlier time after the formation of the first core\footnote{This is a later absolute time, but the first core formation time is difficult to evaluate at this low resolution.} We thus conclude that the "Shannon dissipation" does not play a significant role in the problem. As for now, we yet have to determine the origin of this issue.

\subsubsection{Effect of the Hall resistivity magnitude}

\begin{figure}
\begin{center}
\includegraphics[trim= 2cm 1cm 2cm 1cm, width=0.49\textwidth]{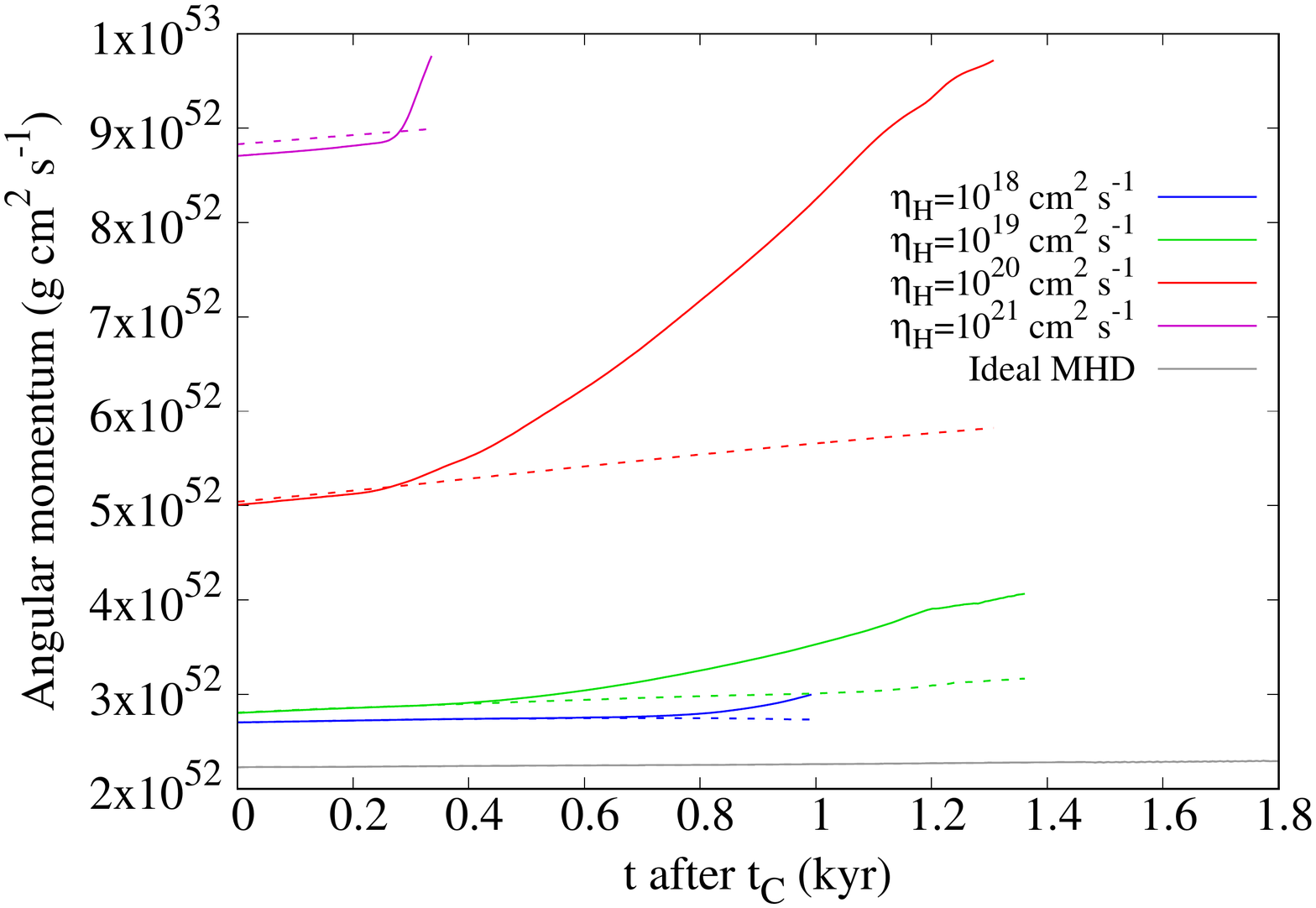}
  \caption{Evolution of the angular momentum from the first Larson core formation for various Hall resistivities.}
\label{etaval}
\end{center}
\end{figure}

We repeated the experiment with various Hall resistivities, from $\eta_\mathrm{H}=10^{18}$ cm$^2$ s$^{-1}$ to $\eta_\mathrm{H}=10^{21}$ cm$^2$ s$^{-1}$, which are relevant values in this context \citep{2016A&A...592A..18M}. Table \ref{table_LC_time} shows the formation time of the first Larson core for the four cases, and the ideal MHD case for comparison. Unsurprisingly, a higher Hall resistivity induces a faster rotation speed, which delays the collapse because of the increasing centrifugal force. Below $10^{19}$ cm$^2$ s$^{-1}$, the induced rotation is weak and barely contributes to the centrifugal support in the isothermal stage.
Figure \ref{etaval} displays the total angular momentum evolution for all the cases. A higher resistivity leads to an earlier and a steeper increase of the angular momentum, and the ideal case shows perfect conservation.

\begin{table}
  \caption{Larson core formation time for different Hall resistivities.}
\label{table_LC_time}
\centering
\begin{tabular}{ll}
\hline\hline
  $\eta_\mathrm{H}$ (cm$^2$ s$^{-1}$) & $t_\mathrm{c}$ (kyr)\\
\hline
  $0$ (ideal case) & $33.550$\\
  $10^{18}$ & $33.130$ \\
  $10^{19}$ & $33.146$ \\
  $10^{20}$ & $33.287$ \\
  $10^{21}$ & $34.352$ \\
\hline
\end{tabular} 
\end{table}

\subsubsection{Effect of the Ohmic and ambipolar diffusions}

\begin{figure}
\begin{center}
\includegraphics[trim= 2cm 1cm 2cm 1cm, width=0.49\textwidth]{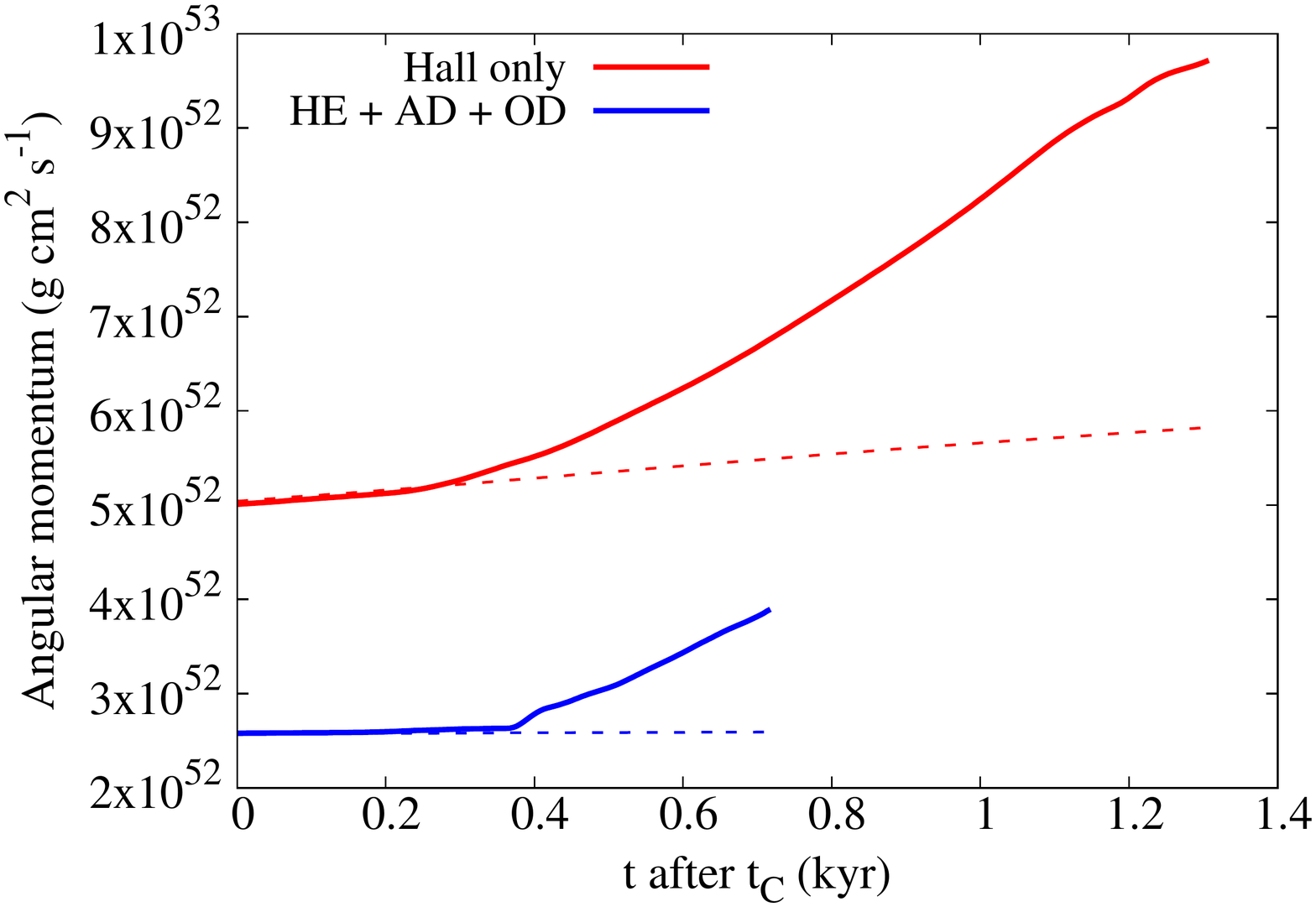}
  \caption{Angular momentum evolution in the reference case (red curves) and with the three non-ideal MHD effects (blue curves).}
\label{ohmad}
\end{center}
\end{figure}

We now aim to include ambipolar and Ohmic diffusions with the Hall term to see whether the other two non-ideal MHD effect can change the angular momentum increase. For the implementation of the ambipolar and Ohmic diffusions in {\ttfamily RAMSES}, we refer the reader to \cite{masson_nimhd}. Figure \ref{ohmad} shows the evolution of the angular momentum for $\eta_\Omega = \eta_\mathrm{H}=\eta_\mathrm{AD} = 10^{20}$ cm$^2$ s$^{-1}$. The divergence occurs $\sim 400$ years after the first Larson core formation, which is a short $100$ years delay compared with the reference case, and with a sharper initial increase. While the ambipolar and Ohmic diffusions influence the dynamics of the collapse, their respective electric field, $\mathbf{E}_\mathrm{AD} \parallel (\mathbf{J} \times \mathbf{B})\times \mathbf{B}$ and $\mathbf{E}_\Omega \parallel \mathbf{J}$ are both perpendicular to the Hall electric field $\mathbf{E}_\mathrm{H} \parallel \mathbf{J}\times\mathbf{B}$, so they do not counter-act the twisting of the field lines induced by the Hall effect. However, the ambipolar and Ohmic diffusions also dissipate and redistribute the magnetic field, effectively weakening the Hall effect in the core, which could be at the origin of the short delay. Nonetheless, our tests show that unless the Hall resistivity is several orders of magnitudes lower than the ambipolar and Ohmic resistivities, the two later effects do not naturally damp the short whistler waves faster than the numerical scheme.

\section{Conclusions}\label{sect_ccl}

We have implemented the Hall effect in the AMR code {\ttfamily RAMSES} to study its impact on star formation and circumstellar disks. The method relies on the modification of the 2D HLL Riemann solver, at the heart of the constrained transport scheme. Our scheme shows a second order convergence in space and the whistler waves are propagated at the correct frequencies. However, the truncation errors leads to a strong numerical dissipation of short whistler waves, as the fourth power of the wave number, and the AMR method seems inefficient to reduce this dissipation. 

During the collapse of a dense core, the Hall effect generates the rotation of the fluid, and thus plays a significant role in the regulation of the angular momentum. While the gas in the mid-plane is set into rotation (in a direction that depends on the sign of $\eta_\mathrm{H}$ and the direction of the magnetic field), counter-rotating envelopes develop on each side of the mid-plane by conservation of the angular momentum. This behavior has been observed in previous studies and was therefore expected.

Roughly 300 years after the first Larson core formation, the angular momentum increases rapidly in the simulation box, meaning that it is no longer conserved. We show that contrarily to the interpretation of \citet{2011ApJ...733...54K} of a similar phenomenon, this problem is not related to our boundary conditions. It emerges for every one of our simulations including the Hall effect. Since this divergence begins shortly after the formation of the first Larson core, it may be correlated to the development of the accretion shock, with only few cells describing the sharp transition. However, to date, we have no clear-cut answer concerning the origin of this problem and various leads will be explored in a further work.

Nonetheless, angular momentum conservation is a key issue in numerical simulations exploring the formation of protostellar disks. There is however no mention of this specific problem in previous studies, most of them using different numerical methods. In order to unambiguously assess the validity of disk formation in such simulations, this issue must at the very least be discussed in detail. A future work will feature additional tests to continue investigating this problem.

\begin{acknowledgements}
  We greatly thank Geoffroy Lesur, Patrick Hennebelle, Yusuke Tsukamoto and Kengo Tomida for the interesting ideas and discussions. We also thank the anonymous referee for his thorough report and comments which helped to improve the clarity of the manuscript. We acknowledge financial support from "Programme National de Physique Stellaire" (PNPS) of CNRS/INSU, CEA and CNES, France, and from an International Research Fellowship of the Japan Society for the Promotion of Science. This work was granted access to the HPC resources of CINES (Occigen) under the allocation DARI A0020407247 made by GENCI. Computations were also performed at the Common Computing Facility (CCF) of the LABEX Lyon Institute of Origins (ANR-10-LABX-0066).
\end{acknowledgements}

\begin{appendix}

  \section{Analytical solution of the steady-state shock}\label{shockanal}

 In this section, we calculate the analytical profile of the shock presented in section \ref{shocktest}.
We first cancel all derivatives except $\frac{\partial}{\partial x}$ in Equations \eqref{testmass}-\eqref{testenergy}. This yields

\begin{align}
  & \frac{d \rho u_x}{dx} = 0, \label{shock1}\\
& \frac{d}{dx} \left(\begin{array}{c}  u_xB_y - u_yB_x \\ u_xB_z - u_zB_x \end{array}\right)  =  \nonumber \\
& \frac{d}{dx} \left(\begin{array}{c}  \left[\eta_\Omega + \eta_\mathrm{AD}\left(1-\frac{B_z^2}{B^2}\right)\right]\frac{dB_y}{dx}  + \left[\eta_\mathrm{H}\frac{B_x}{B}+\eta_\mathrm{AD} \frac{B_yB_z}{B^2}\right] \frac{dB_z}{dx}   \\ \left[-\eta_\mathrm{H}\frac{B_x}{B}+\eta_\mathrm{AD} \frac{B_yB_z}{B^2}\right] \frac{dB_y}{dx}   +   \left[\eta_\Omega + \eta_\mathrm{AD}\left(1-\frac{B_y^2}{B^2}\right)\right]\frac{dB_z}{dx} \end{array}\right), \label{steady_induc}    \\
& \frac{d((\rho u_x)u_x + \rho c_\mathrm{s}^2 + B^2/2)}{dx} = 0, \label{shock2}\\
& \frac{d((\rho u_x)u_y - B_xB_y)}{dx} = 0, \label{shock3}\\
  & \frac{d((\rho u_x)u_z - B_xB_z)}{dx} = 0, \label{shock4}\\
 & \frac{dB_x}{dx} = 0. \label{shock5}
\end{align}

The last Equation \eqref{shock5} is needed to satisfy the divergence-free condition. Integrating Equation \eqref{steady_induc} gives two coupled differential equations in $B_y$ and $B_z$. Isolating the derivatives yields
\begin{align}
 \frac{\partial B_y}{dx} = &\frac{M_\mathrm{1}R_\mathrm{22}-M_\mathrm{2}R_\mathrm{12}}{R_\mathrm{11}R_\mathrm{22} - R_\mathrm{21}R_\mathrm{12}}, \label{dbx} \\
 \frac{\partial B_z}{dx} = &\frac{M_\mathrm{2}R_\mathrm{11}-M_\mathrm{1}R_\mathrm{21}}{R_\mathrm{11}R_\mathrm{22} - R_\mathrm{21}R_\mathrm{12}}, \label{dby} 
\end{align}

with
\begin{equation}
M = \left( \begin{array}{c} u_xB_y-u_yB_x-u_\mathrm{y,0}B_x+u_\mathrm{x,0}B_\mathrm{y,0} \\ u_xB_z-u_zB_x-u_\mathrm{z,0}B_x+u_\mathrm{x,0}B_\mathrm{z,0} \end{array}\right)
\end{equation}

and
\begin{equation}
R = \left( \begin{array}{ll}  \eta_\Omega + \eta_\mathrm{AD}(1-\frac{B_z^2}{B^2})  & \eta_\mathrm{H}\frac{B_x}{B} + \eta_\mathrm{AD}B_yB_z \\  - \eta_\mathrm{H}\frac{B_x}{B} + \eta_\mathrm{AD}B_yB_z & \eta_\Omega + \eta_\mathrm{AD}(1-\frac{B_y^2}{B^2}).  \end{array}\right)
\end{equation}

 Equations \eqref{dbx} and \eqref{dby} are integrated using a Runge-Kutta scheme. We then used the constant quantities $Q = \rho u_x$ (Equation \ref{shock1}) and $K=Qu_x + \frac{Q}{u_x}c_\mathrm{s}^2 + \frac{B^2}{2}$ (Equation \ref{shock2}) to compute the density and velocity profiles.

\begin{align}
& u_x = \frac{1}{2Q}\left(K-\frac{B^2}{2} - \sqrt{\left(K-\frac{B^2}{2}\right)^2 - 4Q^2c_\mathrm{s}^2}\right),  \\
& u_y = u_{y,0} + \frac{B_x(B_y-B_{y,0})}{Q}, \\
& u_z = u_{z,0} + \frac{B_x(B_z-B_{z,0})}{Q}, \\
& \rho = \frac{Q}{u_x}.
\end{align}

  \section{Angular momentum issue for different numerical parameters}\label{app_slopebound}

\begin{figure}
\begin{center}
\includegraphics[trim= 2cm 1cm 1cm 1cm, width=0.49\textwidth]{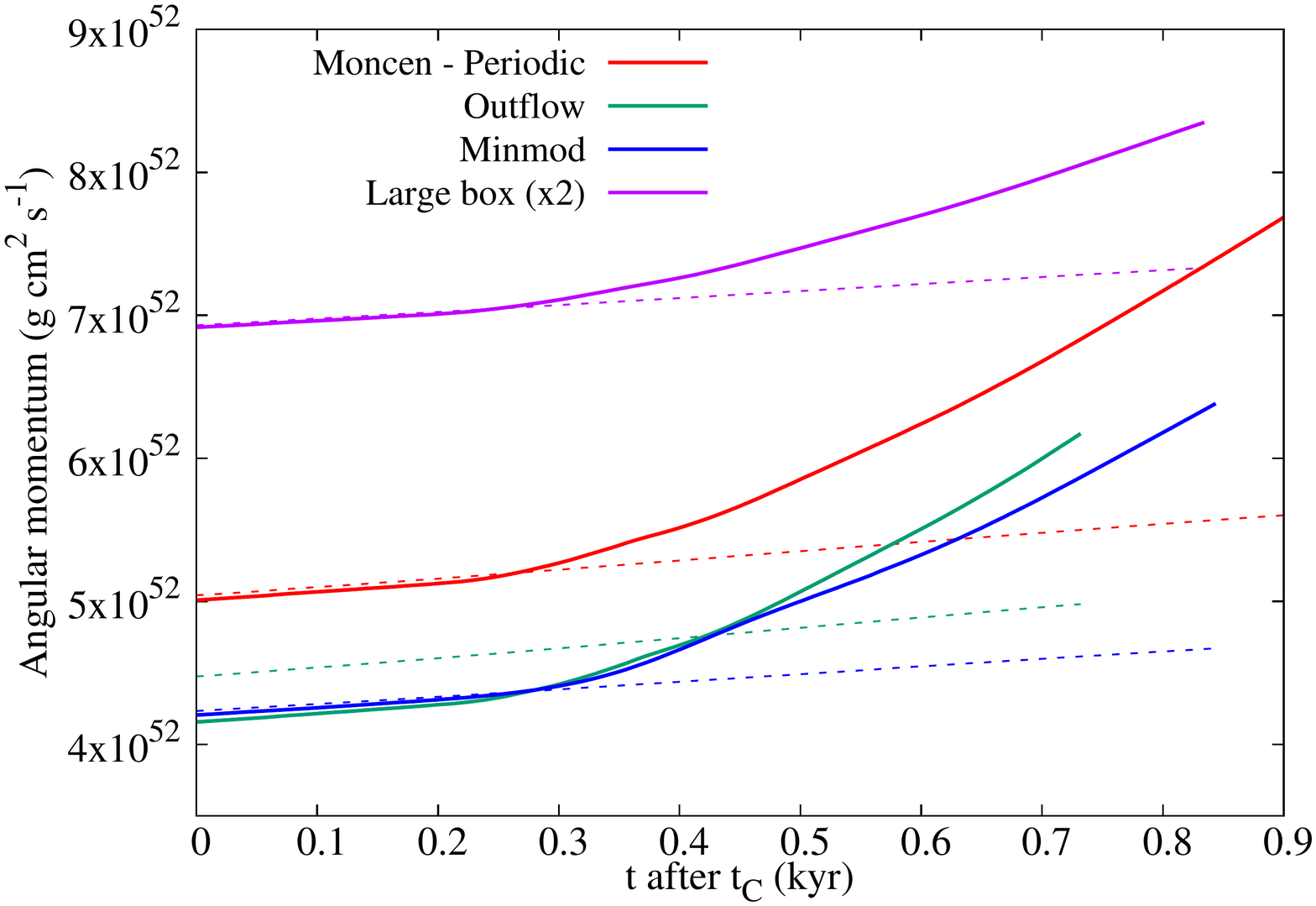}
  \caption{Positive $L_z^+$ (solid lines) and negative $L_z^-$ (dashed lines) angular momentum evolution after the formation of the first Larson core for several variations of the reference case. Red curves: reference case (generalized moncen slope limiter and periodic boundary conditions), green curves: outflow boundary conditions, blue curves: minmod slope limiter, purple curves: simulation box twice as large. The value of the angular momentum has been divided by two for the later to the improve readability of the graph.}
\label{slopebound}
\end{center}
\end{figure}

 In this section, we present the evolution of angular momentum for three variations of the numerical parameters compared to the reference case (moncen and periodic).
 The angular momentum is displayed in Figure \ref{slopebound}, with the reference case in red. In all cases, the divergence stars $300$ years after the formation of the first Larson core and is roughly the same.

 With the minmod slope limiter (blue curves), the angular momentum evolution is roughly the same. In this case, the order of the scheme is first order instead of second, which indicates that the truncation errors do not play a significant role in this problem.

 Outflow boundary conditions (green curves) show a larger initial difference between $L_z^+$ and $L_z^-$ because a fraction of the gas leaves the simulation box. The angular momentum divergence still presents the same evolution as in other cases.

 The purple curves represent a larger box (with a length of 8 times the dense core radius instead of 4). Both angular momentum components have a higher value due to the larger amount of available mass in the box. The divergence starts again $300$ years after $t_\mathrm{C}$ with a comparable increase. These last two cases show that the divergence does not seem related to boundary conditions.

\end{appendix}

\bibliographystyle{aa}
\bibliography{MaBiblio}

\end{document}